\documentclass[12pt,preprint]{aastex}
\usepackage{hyperref}

\shorttitle{CME Sun-to-Earth Propagation}

\shortauthors{Liu et al.}

\begin{document}

\title{On Sun-to-Earth Propagation of Coronal Mass Ejections}

\author{Ying D. Liu\altaffilmark{1,2}, Janet G. Luhmann\altaffilmark{2},
No\'{e} Lugaz\altaffilmark{3}, Christian
M\"{o}stl\altaffilmark{4,2}, Jackie A. Davies\altaffilmark{5},
Stuart D. Bale\altaffilmark{2}, and Robert P. Lin\altaffilmark{2,6}}

\altaffiltext{1}{State Key Laboratory of Space Weather, National
Space Science Center, Chinese Academy of Sciences, Beijing, China}

\altaffiltext{2}{Space Sciences Laboratory, University of
California, Berkeley, CA 94720, USA; liuxying@ssl.berkeley.edu}

\altaffiltext{3}{Space Science Center, University of New Hampshire,
Durham, NH 03824, USA}

\altaffiltext{4}{Institute of Physics, University of Graz, Graz,
Austria}

\altaffiltext{5}{Space Science and Technology Department, Rutherford
Appleton Laboratory, Didcot, UK}

\altaffiltext{6}{School of Space Research, Kyung Hee University,
Yongin, Gyeonggi 446-701, Korea}

\begin{abstract}

We investigate how coronal mass ejections (CMEs) propagate through,
and interact with, the inner heliosphere between the Sun and Earth,
a key question in CME research and space weather forecasting. CME
Sun-to-Earth kinematics are constrained by combining wide-angle
heliospheric imaging observations, interplanetary radio type II
bursts and in situ measurements from multiple vantage points. We
select three events for this study, the 2012 January 19, 23, and
March 7 CMEs. Different from previous event studies, this work
attempts to create a general picture for CME Sun-to-Earth
propagation and compare different techniques for determining CME
interplanetary kinematics. Key results are obtained concerning CME
Sun-to-Earth propagation: (1) the Sun-to-Earth propagation of fast
CMEs can be approximately formulated into three phases: an impulsive acceleration,
then a rapid deceleration, and finally a nearly constant speed
propagation (or gradual deceleration); (2) the CMEs studied here are
still accelerating even after the flare maximum, so energy must be
continuously fed into the CME even after the time of the maximum
heating and radiation has elapsed in the corona; (3) the rapid
deceleration, presumably due to interactions with the ambient
medium, mainly occurs over a relatively short time scale following
the acceleration phase; (4) CME-CME interactions seem a common
phenomenon close to solar maximum. Our comparison between different
techniques (and data sets) has important implications for CME
observations and their interpretations: (1) for the current cases,
triangulation assuming a compact CME geometry is more
reliable than triangulation assuming a spherical front
attached to the Sun for distances below 50-70 solar radii from the Sun, but beyond
about 100 solar radii we would trust the latter more; (2) a proper
treatment of CME geometry must be performed in determining CME
Sun-to-Earth kinematics, especially when the CME propagation
direction is far away from the observer; (3) our approach to
comparing wide-angle heliospheric imaging observations with
interplanetary radio type II bursts provides a novel tool in
investigating CME propagation characteristics. Future CME
observations and space weather forecasting are discussed based on 
these results.

\end{abstract}

\keywords{shock waves --- solar-terrestrial relations --- solar wind
--- Sun: coronal mass ejections (CMEs) --- Sun: flares
--- Sun: radio radiation}

\section{Introduction}

Coronal mass ejections (CMEs) are large-scale expulsions of plasma
and magnetic field from the solar corona and drivers of major space
weather effects. A key question in CME research and space weather
studies is how CMEs propagate through, and interact with, the inner
heliosphere between the Sun and Earth. The crucial importance of
characterizing CME Sun-to-Earth propagation is evident, because this
will lead to a practical capability in space weather forecasting
which has significant consequences for life and technology on the
Earth and in space. Investigation of CME propagation in the
heliosphere requires coordinated remote sensing observations and in situ
measurements from multiple vantage points, which, however, in turn
demand special techniques for their joint interpretation.

Previous studies suggest that fast CMEs undergo a propagation phase
with a constant or slowly decreasing speed, following the
acceleration phase which ceases near the peak time of the associated
soft X-ray flare \citep[e.g.,][]{zhang01}. These studies are based
on coronagraph observations, whose field of view (FOV) is only out
to 30 solar radii. CME speeds can change significantly beyond the
FOV of coronagraphs due to interactions with the ambient solar wind.
Empirical models of CME interplanetary propagation have been
inferred from coronagraph observations and in situ measurements when
they are in quadrature \citep[e.g.,][]{lindsay99, gopalswamy01a}. To
match the speed derived from coronagraph observations near the Sun
with the in situ speed at 1 AU, \citet{gopalswamy01a} assume that
CMEs decelerate out to 0.76 AU from the Sun, after which they travel
with a constant speed. The deceleration cessation distance could not
be precisely determined due to the large observational gap between
the Sun and 1 AU. It is expected that CMEs experience different
propagation histories according to their initial speeds. A similar
model is proposed using interplanetary radio type II bursts whose
frequency drift is a measure of the speed of CME-driven shocks
\citep{reiner07, liu08a}. Based on a statistical analysis of CMEs
that are associated with interplanetary type II bursts,
\citet{reiner07} suggest that CME deceleration can cease anywhere
from about 0.3 AU to beyond 1 AU. The actual situation of CME
interplanetary propagation also involves interactions with the
highly structured solar wind. Global magnetohydrodynamic (MHD)
models \citep[e.g.,][]{riley03, manchester04, odstrcil04} attempt to
create pictures consistent with observations, but until now
realistic case studies have been limited. As a consequence, the
physics governing CME propagation through the entire inner
heliosphere is not well understood.

Now improved determination of CME kinematics over a large distance
is feasible with multiple views of the Sun-Earth space from the
Solar Terrestrial Relations Observatory \citep[STEREO;][]{kaiser08}.
STEREO comprises two spacecraft, one preceding the Earth (STEREO A)
and the other trailing behind (STEREO B), which move away from the
Earth by about 22.5$^{\circ}$ per year. Each of the STEREO
spacecraft carries an identical imaging suite, the Sun Earth
Connection Coronal and Heliospheric Investigation
\citep[SECCHI;][]{howardra08}, which consists of an EUV imager
(EUVI), two coronagraphs (COR1 and COR2) and two heliospheric
imagers (HI1 and HI2). Combined together these cameras can image a
CME from its birth in the corona all the way to the Earth and
beyond. Aboard STEREO there is also a radio and plasma waves
instrument \citep[SWAVES;][]{bougeret08}, which can probe CME
propagation from liftoff to the Earth by measuring type II bursts
associated with CMEs. It covers frequencies from 2.5 kHz to 16 MHz,
which correspond to a range of distance from about 2 solar radii to
1 AU. STEREO also has several sets of in situ instrumentation, which
provide measurements of the magnetic field, energetic particles and
the bulk solar wind plasma \citep{luhmann08, galvin08}. At L1, Wind
and ACE monitor the near-Earth solar wind conditions, thus adding a
third vantage point for in situ measurements. Wind also carries a
WAVES instrument similar to SWAVES \citep[][]{bougeret95}. Details
of when, where, and how CMEs accelerate/decelerate in interplanetary
space can now be quantified by combining wide-angle heliospheric
imaging observations, interplanetary radio type II bursts and in
situ measurements at 1 AU from these multiple vantage points.

Characterizing CME Sun-to-Earth propagation is a primary scientific
question that STEREO was designed to address and the greater
heliospheric constellation enables. Although studies have been
performed attempting to connect STEREO heliospheric imaging
observations with in situ signatures at 1 AU
\citep[e.g.,][]{davis09, wood09, rouillard09, mostl09, mostl10,
liu10a, liu10b, liu11, liu12, harrison12, temmer12, lugaz12,
webb13}, there is much to learn on how to properly interpret these
combined data sets and their information on CME propagation from the
Sun to the Earth. A major limitation is the inherent difficulties
involved in interpreting the heliospheric imaging observations. A
popular approach is to fit CME tracks in time-elongation maps
reconstructed from the wide-angle imaging observations by assuming a
kinematic model \citep[e.g.,][]{sheeley08, rouillard08, lugaz10a,
mostl11, davies12}. CMEs are assumed to travel along a constant
direction at a constant speed in these models. A recent statistical
analysis indicates that these track fitting techniques, while
useful, give rise to a dependence between the free parameters
\citep[][see their Figure~3]{davies12}. Clearly, the degeneracy
between the free parameters cannot be broken in a fitting technique,
so the solutions are not unique. \citet{liu10a, liu10b} have
developed a geometric triangulation technique using the stereoscopic
imaging observations from STEREO. The uniqueness of this method is
that it has no free parameters and can determine CME kinematics as a
function of distance, which provides an unprecedented opportunity to
probe CME propagation and interaction with the heliosphere. The
technique has had success in tracking CMEs and connecting imaging
observations with in situ signatures for various CMEs and spacecraft
longitudinal separations \citep{liu10a, liu10b, liu11, liu12,
mostl10, harrison12, temmer12}. It assumes that the two spacecraft
observe generally the same part of a CME, which, however, may not be
always true, especially for extreme cases like the events studied here
(very wide and back-sided for STEREO). \citet{lugaz10b}
and \citet{liu10b} realize that the triangulation concept can be
applied with a harmonic mean (HM) approach, which assumes that CMEs 
have a spherical front attached to the Sun and what is seen by a spacecraft
is the segment tangent to the line of sight \citep{lugaz09}. The HM
approximation is intended for wide CMEs only. Also note that CMEs
can be significantly distorted by ambient coronal and solar wind
structures \citep[e.g.,][]{riley04, liu06, liu08b, liu09b}. It would
be interesting to compare results from the two triangulation
techniques, which is one of the purposes of this paper. It should be
kept in mind that the present CMEs are extreme cases, i.e., very
wide and back-sided events for STEREO. Any discussions and
implications on the advantages and disadvantages of the two
triangulation methods based on the results should be taken with
caution.

This paper has a threefold aim: (1) to constrain CME Sun-to-Earth
propagation using coordinated heliospheric imaging, radio and in
situ observations; (2) to compare different techniques with which to
determine CME kinematics in interplanetary space; (3) to investigate
crucial physical processes governing CME propagation and interaction
with the heliosphere. This work requires heliospheric imaging
observations from both STEREO A and B, interplanetary radio type II
bursts of a sufficiently long duration (so that comparison with
wide-angle heliospheric imaging observations is meaningful), and in
situ signatures at 1 AU. Due to the last long solar minimum,
long-duration interplanetary type II bursts have only been available
since 2011
(\url{http://lep694.gsfc.nasa.gov/waves/data_products.html}). By
this time, the two STEREO spacecraft were already behind the Sun, so
the Earth-directed CMEs after 2011 (February) are essentially
back-sided events for STEREO. From their vantage points behind the
Sun, the two spacecraft are likely observing different parts of an
Earth-directed CME when it is at large distances. In particular,
CMEs that are associated with a long-duration interplanetary type II
burst are usually wide events. This is why triangulation with the HM
approximation is likely to be necessary. Observations and
methodology are described in Section 2. We present details on case
studies in Section 3. The results are summarized and discussed in
Section 4. To the best of our knowledge, this is the first study
that attempts to create a general picture of CME Sun-to-Earth
propagation combining wide-angle heliospheric imaging observations,
interplanetary type II bursts and in situ measurements at 1 AU. Our
findings on CME Sun-to-Earth propagation provide a test bed or
benchmark for CME Sun-to-Earth research and space weather
prediction. New insights obtained from this work will improve
observational strategies and analysis tools. Specifically, this work
evaluates a triangulation concept for future CME observations and
space weather forecasting, a key strategic study to prepare for
future missions at L4/L5.

\section{Observations and Methodology}

We pick three events for this study, i.e., the 2012 January 19, 23
and March 7 CMEs. Each of these events has wide-angle imaging
coverage from both STEREO A and B, a long-duration interplanetary
type II burst, and in situ signatures near the Earth. These are
extreme cases, i.e., wide and back-sided for STEREO. Below we
describe the observations and methodology used for the joint
interpretation of the different data sets.

\subsection{Geometric Triangulation of Imaging Observations}

Of particular interest for this study are the heliospheric imagers
(HI1 and HI2) on board STEREO. They can observe CMEs out to the vicinity 
of the Earth and beyond by using sufficient baffling to eliminate
stray light \citep{harrison08, eyles09}. The combined effects of
projection, Thomson scattering and CME geometry, however, form a
major challenge in determining CME kinematics from the wide-angle
imaging observations. The geometric triangulation technique
developed by \citet{liu10a, liu10b} can determine both propagation
direction and radial distance of CMEs from STEREO imaging
observations, assuming a fixed $\beta$ (F$\beta$) geometry, i.e., a
relatively compact structure simultaneously seen by the two
spacecraft. (The F$\beta$ approximation was initially used for track fitting, which
assumes a compact structure moving along a fixed radial direction
\citep{sheeley99, kahler07, liu10b}. This is where the term
``fixed'' comes from. When the F$\beta$ formula is adopted for
geometric triangulation, the assumption of fixed radial trajectory
is abolished.) Hereinafter we call the technique triangulation with the F$\beta$
approximation. \citet{liu10a, liu10b} describe the mathematical
formulas and detailed procedures for applying this technique. The
F$\beta$ triangulation gives good results in analysis of
Earth-oriented CMEs when the two spacecraft are before the Sun
\citep{liu10a, liu10b, liu11, liu12, mostl10, harrison12, temmer12}.
Now we evaluate its use on those wide and back-sided events. 

We also apply the triangulation concept under a harmonic mean (HM)
approximation, which takes into account the effect of CME geometry.
It assumes that a CME has a spherical front attached to the Sun, and the
two spacecraft observe different plasma parcels that lie along the
tangent to their line of sight \citep[][]{lugaz09}. Hereinafter we
refer to it as triangulation with the HM approximation. Both of the
triangulation techniques have no free parameters, and can determine
CME kinematics as a function of distance from the Sun continuously
out to 1 AU. These advantages make the techniques a powerful tool
for investigating CME Sun-to-Earth propagation and forecasting space
weather. The readers are directed to \citet{liu10b} and
\citet{lugaz10b} for more detailed discussions of these two
triangulation techniques. In this work, we apply the two methods to
the imaging observations of COR2, HI1 and HI2. COR1 images are also
examined but not included here given its small FOV. Therefore, the
resulting CME kinematics do not include the initiation phase of
CMEs. We will compare the two methods using the current cases and
see how they work as we move from small to large distances. We will
also investigate how their results compare with other data sets.

\subsection{Frequency Drift of Radio Type II Bursts}

Interplanetary radio type II bursts associated with CMEs provide
further constraints on CME Sun-to-Earth propagation. Type II bursts,
typically drifting downward in frequency, are plasma radiation near
the local plasma frequency and/or its second harmonic produced by a
shock \citep[e.g.,][]{nelson85, cane87}. The frequency drift results
from the decrease of the plasma density as the shock moves away from
the Sun, so the shock propagation distance can be inverted from the
type II burst by assuming a density model \citep[e.g.,][]{reiner07,
liu08a, gonzalez09}. In this study, we use a density model which
covers a distance range from about 1.8 solar radii to 1 AU
\citep[][hereinafter referred to as the Leblanc density
model]{leblanc98}. The Leblanc density model is scaled by a density
of 6.5 cm$^{-3}$ at 1 AU, the average value of Wind measurements
over a few months around the CME times. The same density model is
applied to all the three events.

Note that the density model describes an average radial variation of
the ambient solar wind density, so we do not expect that it fits the
band of a type II burst all the times. Recent work has shown that
the source position of a type II burst at a particular frequency can
be triangulated without using a density model \citep{juan12}, but
that technique requires a strong radio emission seen by more than
one spacecraft. The current cases are all back-sided events for
STEREO, and strong type II bursts are only observed by Wind. This
work is intended to compare wide-angle heliospheric imaging
observations with long-duration interplanetary type II bursts. In
doing this we explicitly assume that the distances inverted from a
type II burst, which essentially represent shock kinematics, do not
deviate much from the distances of the leading edge seen in white
light. The geometric triangulation described above is applied to the
leading feature in imaging observations, which could be a
high-density structure in the sheath between the shock and ejecta or
the shock itself as revealed by previous studies \citep{liu11,
liu12}.

\subsection{Comparison with In Situ Measurements}

The CME kinematics inferred from imaging observations and type II
bursts will be compared with in situ measurements at Wind, another
step in our analysis. CMEs are called interplanetary CMEs (ICMEs)
when they move into the solar wind. In situ measurements can give
local plasma and magnetic field properties along a one-dimensional
cut when ICMEs encounter spacecraft. Signatures used to identify
ICMEs from solar wind measurements include depressed proton
temperatures, enhanced helium/proton density ratio, smooth magnetic
fields, and rotation of the field. In this paper we use the term
``CMEs" for events observed in images and ``ICMEs" for ejecta
identified from in situ measurements. Our comparison with in situ
data mainly focuses on CME speed and arrival time at the Earth. The
connections between imaging observations, type II bursts and in situ
signatures can also indicate the nature of structures observed in
images, e.g., shocks and flux ropes \citep[][]{liu10a, liu10b,
liu11, liu12, deforest11}.

\section{Case Studies}

Figure~1 shows the trajectories of the CMEs of interest as well as
the configuration of the planets and spacecraft in the ecliptic
plane. STEREO A and B are separated by about 221.1$^{\circ}$ in
longitude (the angle bracketing the Earth) on 2012 January 19, about
221.7$^{\circ}$ on January 23 and about 227.1$^{\circ}$ on March 7,
respectively. Triangulation with the HM approximation gives a
propagation angle with respect to the Sun-Earth line roughly twice
of that from triangulation with the F$\beta$ approximation, as
previously indicated by \citet{lugaz10a}. During the 2012 March 7
CME, Mars was at 1.66 AU from the Sun and almost radially aligned
with the Earth, so the March 7 CME was likely to impact both the
Earth and Mars. Enhanced fluxes of energetic particles were indeed
observed by Curiosity, when it was traveling to Mars inside
the Mars Science Laboratory spacecraft
(\url{http://science.nasa.gov/science-news/science-at-nasa/2012/02aug_rad2/}).

\subsection{2012 January 19 Event}

The 2012 January 19 CME originated from NOAA AR 11402
(N32$^{\circ}$E22$^{\circ}$), and is associated with a long-duration
M3.2 flare which peaked around 16:05 UT on January 19. Figure~2
(left) displays two synoptic views of the CME from STEREO A and B.
Only shown are images from COR2, HI1 and HI2. COR2 has a
0.7$^{\circ}$ - 4$^{\circ}$ FOV around the Sun. HI1 has a
20$^{\circ}$ square FOV centered at 14$^{\circ}$ elongation from the
center of the Sun while HI2 has a 70$^{\circ}$ FOV centered at
53.7$^{\circ}$. The CME is launched from the Sun at about 13:55 UT
on January 19 and has a peak speed of about 1200 km s$^{-1}$ (see
Figure~1). Deflections of remote coronal structures are observed in
COR2 images, which suggest the existence of the CME-driven shock
\citep[e.g.,][]{vourlidas03, liu09a}. The CME looks quite different
for STEREO A and B, especially in HI1 and HI2, which may indicate
the significant effect on the CME of the medium through which it is
propagating. Figure~2 (right) shows the time-elongation maps, which
are produced by stacking the running difference intensities of COR2,
HI1 and HI2 within a slit around the ecliptic plane
\citep[e.g.,][]{sheeley08, davies09, liu10a}. Multiple tracks are
observed on January 18 - 19, so CME-CME interactions are likely
present. In particular, a CME from January 18 and another one that
occurred a few hours earlier on January 19 are visible to both
STEREO A and B, and could have effect on the interplanetary
propagation of the CME of interest. In STEREO B observations the CME
reaches the elongation of the Earth earlier than in STEREO A images,
so it must propagate east of the Sun-Earth line.

We apply the two triangulation techniques to the leading feature of
the CME in imaging observations, and the resulting CME kinematics in
the ecliptic plane are displayed in Figure~3. Both of the methods
give negative propagation angles (east of the Sun-Earth line),
consistent with the solar source location and the time-elongation
maps. The propagation angle from triangulation with the F$\beta$
approximation starts from about the solar source longitude, while
the angle from triangulation with the HM approximation keeps about
twice of the former and finally approaches the solar source
longitude. Their trends are similar except that the HM approximation
enlarges the variations. The distances from the two methods show no
essential differences below about 70 solar radii, but after about
100 radii the F$\beta$ approximation gives an apparent acceleration.
This late acceleration is also noticed in previous studies
\citep[e.g.,][]{lugaz09, wood09, harrison12}. It is not physically
meaningful, as there is no obvious force responsible for the
acceleration at large distances from the Sun. A similar scenario is
observed in the 2012 January 23 and March 7 CMEs. In the current
cases the late acceleration at large distances is due to the
non-optimal observation situation (from behind the Sun) in
combination with the restrictions of the F$\beta$ geometry (see
section 4.2).

The speed profiles from the two triangulation methods are similar
below about 60 solar radii, except that the peak speed from the HM
approximation is a little higher. Both the speed profiles show an
impulsive acceleration up to about 15 solar radii, then a rapid
deceleration out to about 35 solar radii, and then a roughly
constant value thereafter. The late increase in the speed profile
from the F$\beta$ approximation, again, is considered unmeaningful.
The CME is still accelerating even after the flare maximum, as
indicated by the timing with the GOES X-ray curve. The CME speed
peaks about 2.5 hours after the flare maximum, so energy must be
continuously fed into the CME even after the time of the maximum
heating and radiation has elapsed in the corona. The rapid
deceleration, presumably due to interactions with the ambient
medium, only lasts about 4 hours, a relatively short time scale.
Therefore the maximum drag by, or energy loss to, the ambient medium
takes place at distances not far from the Sun (within 35 solar
radii). Similar results have been alluded to in other cases
\citep[e.g.,][]{liu10a, liu10b, liu11}. The HM approximation can
suppress the late acceleration, but compared with the average solar
wind speed behind the shock observed in situ near the Earth, it
still overestimates the speed by about 250 km s$^{-1}$ (see
Figure~6).

Figure~4 shows the radio dynamic spectrum associated with the CME.
Strong type II emissions are only observed by Wind. They occur at
the fundamental and harmonic plasma frequencies and appear as slowly
drifting features. The type II burst starts from the upper bound (16
MHz) around 15:00 UT on January 19 and extends down to about 150
kHz. We use an approach different from previous studies: the CME
leading-edge distances derived from imaging observations are
converted to frequencies using the Leblanc density model (with a
density of 6.5 cm$^{-3}$ at 1 AU), given that we already have the
distances. The imaging triangulation results are generally
consistent with the type II burst, and there is no essential
difference between the F$\beta$ and HM approximations for the time
period covered by the figure. The weak type II emissions observed by
STEREO A and B, e.g., those at 400 kHz around 19:30 UT on January
19, are also captured by the triangulation results. The consistency
between the imaging results and radio data from three spacecraft
suggests the validity of our triangulation concept. It may also
indicate a relationship between the CME leading feature being
tracked and the type II burst, but a careful study is needed to
reach a definite conclusion. Note that many short-lived type III
like bursts, known as a type III storm \citep[e.g.,][]{fainberg70}, are also observed by Wind.
This type III storm seems to have an overall drift rate similar to
that of the type II burst, which may hold an important clue about
the origin of the type III storm.

The frequencies of the type II burst (assuming the harmonic branch)
are also converted to distances using the Leblanc density model, as
shown in Figure~5. The error bars in the radio data are determined
from the band width of the type II burst. Again, we see a general
consistency between the imaging triangulation results and the type
II burst, and there is no essential difference between the F$\beta$
and HM approximations below 60 solar radii. There appears to be a decrease
in the drift rate of the type II burst after about 02:00 UT on
January 20, while the CME does not seem to have an apparent slowdown
at the same time.

Figure~6 shows an ICME identified from Wind in situ measurements
based on the depressed proton temperature and smooth magnetic field.
The ICME interval is also consistent with the decreased proton beta.
This is the in situ counterpart of the 2012 January 19 CME. The
CME-driven shock (shock1) passed Wind at about 05:28 UT on January
22. The average speed in the sheath region between the shock and
ICME is about 415 km s$^{-1}$, which is about 250 km s$^{-1}$ lower
than predicted by triangulation with the HM approximation (see
Figure~3). The predicted arrival time at the Earth is about 08:10 UT
on January 21 resulting from triangulation with the F$\beta$
approximation and 03:11 UT on January 22 given by the HM
triangulation. We have used $r\cos\beta$ and $v\cos\beta$ for the HM
approach in estimating the arrival time and speed at the Earth,
respectively, where $r$ is the distance from imaging observations,
$v$ the speed, and $\beta$ the propagation angle with respect to the
Sun-Earth line. The much earlier prediction from the F$\beta$
triangulation compared with the observed shock arrival is probably
owing to the unphysical late acceleration at large distances. The HM
triangulation does a very good job in predicting the shock arrival,
although it overestimates the speed considerably. Compared with
the speed near the Sun ($\sim1200$ km s$^{-1}$), the ICME/shock is
significantly decelerated by the structured solar wind. The second
shock (shock2), which is also the trailing boundary of the ICME, is
driven by the 2012 January 23 CME. The shock passed Wind at about
14:38 UT on January 24, and is beginning to overtake the ICME from
behind.

\subsection{2012 January 23 Event}

The 2012 January 23 CME originated from NOAA AR 11402
(N29$^{\circ}$W20$^{\circ}$), the same active region as the 2012
January 19 event, and is associated with a long-duration M8.7 flare
which peaked around 03:59 UT on January 23. Figure~7 (left) displays
two views of the CME from STEREO A and B. The CME is launched from
the Sun at about 03:40 UT on January 23 and has a peak speed of
about 1600 km s$^{-1}$ (see Figure~1). A weak edge ahead of the CME
front is visible in COR2, reminiscent of a shock signature
\citep[e.g.,][]{vourlidas03, liu08a}. Only part of the CME is seen
in HI1 and HI2, indicating the need for a bigger FOV in future CME
heliospheric imaging observations. In HI2 of STEREO A, the CME
leading feature (which could be the shock) appears as a grand wave
sweeping over the inner heliosphere, while the driver can hardly be
seen. This must be part of the evolutionary history of the densities
in the ejecta and sheath \citep[][]{liu11}. Figure~7 (right) shows
the time-elongation maps produced from intensities around the
ecliptic. Again, other tracks are observed around the time of the
CME, so CME-CME interactions are expected. The CME must propagate
west of the Sun-Earth line, since in STEREO A observations it
reaches the Earth's elongation earlier than in STEREO B.

The CME kinematics resulting from the two triangulation techniques
are shown in Figure~8. The two methods give positive propagation
angles (west of the Sun-Earth line), consistent with the solar
source location and the time-elongation maps. However, the
propagation angles from both of the techniques deviate from the
solar source longitude, which may suggest longitudinal deflections
of the CME at its early stage. Again, the trends in the angles are
similar for the F$\beta$ and HM approximations, and the angle from
the HM approach keeps about twice of that from the F$\beta$
approximation. The F$\beta$ approach gives an unphysical
acceleration after about 100 solar radii, which is due to the
non-optimal observation situation (from behind the Sun) and the
F$\beta$ geometry as discussed above. The speed profiles from the
two triangulation methods are similar below about 75 solar radii.
Both the speed profiles show an impulsive acceleration up to about
16 solar radii, and then a rapid deceleration out to about 75 solar
radii. The CME is still accelerating even after the flare maximum.
The maximum speed lags the peak of the GOES X-ray curve by about 2
hours. The rapid deceleration lasts about 10 hours in this case. The
HM approximation overestimates the observed solar wind speed at the
Earth by about 250 km s$^{-1}$ (see Figure~6), although it
suppresses the late acceleration.

Figure~9 displays the radio dynamic spectrum associated with the
2012 January 23 CME. Strong type II emissions are only observed by
Wind. The type II burst starts from the upper bound (16 MHz) around
04:10 UT on January 23, and extends down to about 40 kHz when the
shock arrives at Wind (14:38 UT on January 24). Apparently this is a
Sun-to-Earth type II burst. Weak type II spots are observed by
STEREO A (e.g., those near 200 kHz around 13:00 UT on January 23 and
near 70 kHz around 06:30 UT on January 24), but not STEREO B. This
indicates that the propagation direction is closer to STEREO A than
B, consistent with the triangulation results. The CME leading-edge
distances from imaging observations, after being converted to
frequencies using the Leblanc density model (with a density of 6.5
cm$^{-3}$ at 1 AU), are generally consistent with the type II burst.
Here we assume that the strong type II band observed by Wind is
emitted at the second harmonic of the local plasma frequency. The
weak type II emissions observed by STEREO A are also captured by the
triangulation results, again indicating the validity of our
triangulation concept. At large distances the F$\beta$ approach
appears to fit the type II burst better than HM, which is not what
we expect.

Figure~10 shows the distances derived from the type II burst using
the Leblanc density model, in a comparison with triangulation
analysis. There is essentially no difference between the F$\beta$
and HM approximations below about 70 solar radii. The F$\beta$
approximation is more consistent with the type II burst than HM at
large distances, as we see from Figure~9. It is not clear about the
cause of the late increase in the frequency drift of the type II
burst. Perhaps the shock is propagating into a less dense medium at
large distances, or it is due to complexities in the radio emission
region. The shock driven by the 2012 January 23 CME (shock2) passed
Wind at about 14:38 UT on January 24, as shown in Figure~6, and is
propagating into the ICME associated with the January 19 event. The
ICME preceding the shock does have a depressed proton density at 1
AU compared with the ambient medium. The average speed behind the
shock is about 650 km s$^{-1}$, again about 250 km s$^{-1}$ lower
than predicted by the HM triangulation. Triangulation with the
F$\beta$ approximation gives a predicted arrival time at the Earth
of about 09:42 UT on January 24, while the HM triangulation yields a
predicted arrival time of about 23:06 UT on January 24. Again, we
have used $r\cos\beta$ and $v\cos\beta$ in estimating the arrival
time and speed at the Earth for the HM approach. It is surprising
that the F$\beta$ triangulation does a better job than the HM
approach in predicting the shock arrival. A possible explanation is 
that the shock was propagating through the preceding ICME at 
large distances, which is less dense than the ambient solar wind. This 
might compensate for the the unphysical late acceleration given by the 
F$\beta$ triangulation. The 2012 January 23 CME does not have in situ
signatures at Wind (except the shock). It is likely that the CME
missed Wind, and the shock must have a structure larger than the
ejecta or flux rope.

\subsection{2012 March 7 Event}

March 2012 is a very interesting period, during which a series of
M/X class flares and big CMEs occurred. In particular, NOAA AR 11429
generated an X1.1 flare and a CME of about 1500 km s$^{-1}$ around
03:00 UT on March 5, an X5.4 flare and a CME of more than 2000 km
s$^{-1}$ around 00:15 UT on March 7, an M6.3 flare and a CME of
about 1000 km s$^{-1}$ around 03:30 UT on March 9, an M8.4 flare and
a CME of more than 1000 km s$^{-1}$ around 17:30 UT on March 10, and
an M7.9 flare and a CME of over 1500 km s$^{-1}$ around 17:30 UT on
March 13. Among them the March 7 CME is the largest and most
energetic. There seems another event on March 7, which occurred
about 1 hour later than the first one. It produced an X1.3 flare and
a CME of over 1500 km s$^{-1}$. The second CME is occulted by the
first one in STEREO coronagraph images, as it occurred close in time
with the first one. At least three of the events hit the Earth,
including the March 5, 7 and 10 events (see Figure~15).

The violent corona and heliosphere during the March 7 CME are shown
in Figure~11 (left), as seen from STEREO A and B. NOAA AR 11429 was
at N17$^{\circ}$E21$^{\circ}$ when the CME occurred, and the
associated long-duration X5.4 flare peaked around 00:24 UT on March
7. The CME is launched from the Sun at about 00:15 UT and has a peak
speed of about 2400 km s$^{-1}$ (see Figure~1). A weak edge around
the CME front is observed in COR2, indicative of a shock signature
\citep[e.g.,][]{vourlidas03, liu08a}. The heliosphere observed by
HI1 and HI2 is very turbulent when the series of dramatic CMEs blow
through their FOVs. Figure~11 (right) shows the time-elongation maps
produced from intensities around the ecliptic. We observe multiple
tracks from the maps, some of which cross, which suggests CME-CME
interactions. The CME must propagate east of the Sun-Earth line,
since in STEREO B observations it reaches the Earth's elongation
earlier than in STEREO A.

The kinematics of the CME leading feature resulting from the two
triangulation techniques are shown in Figure~12. Both of the methods
give negative propagation angles (east of the Sun-Earth line),
consistent with the solar source location and the time-elongation
maps. Similar to the January 19 event, the propagation angle from
triangulation with the F$\beta$ approximation starts from about the
solar source longitude. The angle from triangulation with the HM
approximation still keeps about twice of that from the F$\beta$ approximation.
The F$\beta$ approach gives an unphysical acceleration after about 120 solar
radii, which is due to the non-optimal observation situation (from
behind the Sun) and the assumed CME geometry. There is also a slight
acceleration in the HM approach at large distances, which suggests
that the HM approximation is not adequate to suppress the apparent
late acceleration in this case. The speed profiles from the two
triangulation methods are similar below about 70 solar radii. Both
the speed profiles show an impulsive acceleration up to about 15
solar radii, then a rapid deceleration out to about 55 solar radii,
and then a roughly constant value thereafter. The CME is still
accelerating even after the flare maximum. The maximum speed occurs
about 2 hours after the peak of the GOES X-ray curve. The rapid
deceleration lasts about 5 hours in this case. The HM approximation
overestimates the observed solar wind speed at the Earth by about
270 km s$^{-1}$ (see Figure~15).

Figure~13 displays the radio dynamic spectrum associated with the
2012 March 7 CME. Strong type II emissions are only observed by
Wind. The type II burst starts from the upper bound (16 MHz) around
01:15 UT on March 7, and extends down to about 40 kHz by the time
the shock arrives at Wind (10:19 UT on March 8). This is a
Sun-to-Earth type II burst and probably the strongest one in solar
cycle 24 up to the time of this writing. Weak type II spots are
observed by STEREO B, e.g., those near 200 kHz around 10:00 UT on
March 7, while below 1000 kHz no type II emissions are observed by
STEREO A. This indicates that the propagation direction is closer to
STEREO B than A, consistent with the triangulation results. The CME
leading-edge distances from imaging observations, after being
converted to frequencies using the Leblanc density model (with a
density of 6.5 cm$^{-3}$ at 1 AU), are generally consistent with the
type II burst. We assume that the type II band we are comparing is
emitted at the second harmonic of the local plasma frequency. The
imaging triangulation results also capture the weak type II
emissions observed by STEREO A and B, which again indicates the
validity of our triangulation concept. At large distances the HM
approach fits the type II burst better than F$\beta$. A type III
storm, whose overall drift rate is similar to that of the type II
burst, is also observed by Wind (above the type II burst).

Figure~14 shows the comparison between the distances derived from
the type II burst using the Leblanc density model and triangulation
analysis. There is no essential difference between the F$\beta$ and
HM approximations below about 50 solar radii. The HM approximation
is more consistent with the type II burst than F$\beta$ after March
7, as we see from Figure~13. The type II burst suggests a quick
deceleration on March 8, but the large error bars (from the band
width of the type II burst) do not allow determination of how much
the shock is slowed down. Irregularities in the type II burst, which
come from the non-uniformity of the ambient density, may contribute
to the apparent deceleration.

The in situ signatures at Wind are shown in Figure~15. Two ICMEs are
identified during the time period using the depressed proton
temperature and smooth magnetic field. The first one (ICME1) and its
shock (S1) are produced by the March 7 CME, and the second (ICME2)
and its shock (S2) by the March 10 event. A weak shock (S0)
preceding S1 seems driven by a CME from March 5, but no driver
signatures are observed (except the shock). Interactions between the
March 5 and 7 CMEs are likely to have occurred near 1 AU. The ICMEs,
shocks and possible interactions between them caused a long duration
(from March 7 to March 19) of significant negative values of the
D$_{\rm st}$ index. In particular, ICME1 (and S1) produced a major
geomagnetic storm with a minimum D$_{\rm st}$ of $-143$ nT. The
shock driven by the March 7 CME (S1) passed Wind at about 10:19 UT
on March 8. The average speed in the sheath behind S1 is about 680
km s$^{-1}$, about 270 km s$^{-1}$ lower than predicted by the HM
approximation. Triangulations with the F$\beta$ and HM
approximations give a predicted arrival time at the Earth of about
03:27 UT and 17:55 UT on March 8, respectively. Compared with the
actual shock arrival time (10:19 UT), these predictions appear
equally good, but the F$\beta$ prediction is earlier than the
observed shock arrival and the HM one lies in the sheath. This
raises a question in space weather studies (and forecasting as
well): what in situ signatures should we compare with, the shock or
the leading edge of the flux rope? Previous studies attempting to
connect imaging observations with in situ measurements indicate that
what is being tracked is often a high-density structure in the
sheath or the shock itself \citep{liu11, liu12}. The ICME/shock is
again significantly decelerated by the structured solar wind,
compared with the speed near the Sun ($\sim2400$ km s$^{-1}$).

\section{Conclusions and Discussion}

To the best of our knowledge, this is the first comprehensive study
that attempts to generalize the whole Sun-to-Earth propagation of
CMEs combining heliospheric imaging observations, interplanetary
radio type II bursts and in situ measurements from multiple vantage
points. We summarize and discuss our results as follows, based on
the case studies of the 2012 January 19, 23 and March 7 CMEs.

\subsection{Formulating CME Sun-to-Earth Propagation}

Key findings are revealed on how CMEs propagate through, and
interact with, the entire inner heliosphere between the Sun and
Earth. Below we formulate CME Sun-to-Earth propagation. Any
theory/model of CME Sun-to-Earth propagation and space weather
forecasting should be guided and constrained by the results presented here.

1. The Sun-to-Earth propagation of fast CMEs can be approximately described by
three phases: an impulsive acceleration, then a rapid deceleration,
and finally a nearly constant speed propagation (or gradual
deceleration). The initiation phase in the low corona is not
included in this study. We alluded to similar results in our
triangulation analysis of the 2008 December 12 CME \citep{liu10a,
liu10b} and the 2010 April 3 CME \citep{liu11}. It should be
stressed that we are looking at CME propagation in the whole
Sun-Earth space, not just within 20-30 solar radii as in some
previous studies \citep[e.g.,][]{sheeley99, zhang01}. The speed of fast CMEs
during the gradual deceleration phase is expected to approach that
of the ambient solar wind. This should be a long process (out to
several tens of AU from the Sun) as the energy is slowly dissipated
into the ambient medium \citep[e.g.,][]{richardson05, liu08a}. The
three phases hold only for fast events. CMEs slower than the ambient
solar wind are expected to be accelerated by the solar wind 
\citep[e.g.,][and references therein]{sheeley99}.

2. The CMEs studied here are still accelerating even after the flare
maximum, so energy must be continuously fed into the CME even after
the time of the maximum heating and radiation has elapsed in the
corona. \citet{zhang01} suggest that CME acceleration occurs
simultaneously with the flare impulsive phase and ceases near the
peak of the GOES X-ray curve. Here we find further acceleration post
the flare maximum with a time scale of about 2 hours. This is not
uncommon \citep[e.g.,][]{cheng10, zhao10, liu11}. A statistical
analysis indicates that the post-flare-maximum acceleration tends to
occur in cases associated with a long-duration flare
\citep{cheng10}. The time scale of CME further acceleration after the flare 
maximum may depend on the balance between energy injection from 
the Sun and drag by the ambient medium.

3. The rapid deceleration, presumably due to interactions with the
ambient medium, mainly occurs over a relatively short time scale
following the acceleration phase. The maximum drag by, or energy
loss to, the ambient medium takes place at distances not far from
the Sun. The rapid deceleration lasts about 5-10 hours and stops
around 40-80 solar radii from the Sun for the current cases. The
short cessation distance of CME deceleration is a surprising finding
compared with previous indirect inferences \citep{gopalswamy01a,
reiner07}. While drag by the ambient medium may play an important
role in decelerating CMEs, we expect that a significant portion of
the energy goes into energetic particles through shock acceleration
during the impulsive acceleration and rapid deceleration phases.
\citet{mewaldt08} find that the total energy content of energetic
particles can be 10\% or more of the CME kinetic energy. Obviously
the finding of CME rapid deceleration is of importance
for both space weather forecasting and particle acceleration.

4. CME-CME interactions seem a common phenomenon close to solar
maximum. We see CME-CME interaction signatures in both remote
sensing and in situ observations for the present cases. CME-CME
interactions are expected to be a frequent phenomenon in the
Sun-Earth space near solar maximum. The rationale is that the
typical transit time of CMEs from the Sun to 1 AU is a few days,
much longer than the time within which multiple CMEs occur at solar
maximum. Interactions between CMEs are of importance for both 
space weather studies and basic plasma physics: they can produce 
or enhance southward magnetic fields, modify shock structure, 
particle acceleration and transport, and give rise to significant energy 
and momentum transfer between the interacting systems via magnetic 
reconnection \citep[e.g.,][]{gopalswamy01b, gopalswamy02, burlaga02, 
richardson03, wang03, farrugia04, liu12}. This crucial importance and 
prevalence near solar maximum call attention to CME-CME interactions 
in space weather studies and forecasting.

\subsection{Implications for CME Observations and Interpretations}

Our comparison between different techniques (and data sets) has
important implications for CME observations and their
interpretations. The following insights obtained from this work will
help improve analysis tools and observational strategies.

1. For the current cases, triangulation with the F$\beta$
approximation is more reliable than triangulation with the HM
approximation below 50-70 solar radii from the Sun, but beyond about
100 solar radii we would trust the HM triangulation more. They show
no essential differences in terms of radial distances and speeds
below 50-70 solar radii, but the propagation angle from the HM
approximation is generally twice of that from the F$\beta$ approach.
The F$\beta$ propagation angle is more consistent with the solar
source longitude than the HM angle below roughly the same distance
(i.e., 50-70 solar radii), whereas further out the opposite seems
true. As suggested by the combined solar active region longitudes,
F$\beta$ propagation angles in the lower inner heliosphere and HM
angles in the mid inner heliosphere, there is no indication of
strong longitudinal deflections (several tens of degrees) for the
current cases. Most of the large variations in the CME propagation
angles seem due to the limitations of the methods. The F$\beta$
approximation gives an unphysical acceleration after 100-120 solar
radii. The HM approximation can reduce the late acceleration to some
extent, but it still overestimates the observed speed at the Earth
by about 250 km s$^{-1}$ for the present three cases. Note that the
cutoff distances may vary depending on the size of CMEs and the
longitudinal separation between the two STEREO spacecraft. For
Earth-directed CMEs that are not large and occurred when the two
spacecraft were before the Sun, we obtain satisfactory results using
the F$\beta$ triangulation \citep{liu10a, liu10b, liu11, liu12,
mostl10, harrison12, temmer12}.

It is worth pointing out that the arrival time predictions by
triangulations with the F$\beta$ and HM approximations are more or
less acceptable (except the 2012 January 19 event for the F$\beta$
approach), even though the CME speed at the Earth can be
overestimated considerably. A possible explanation is that, for
fast CMEs like the present ones, it is relatively easy to make a
``good" prediction of the arrival time because the Sun-to-Earth transit
time is short. That could also arise from the shape of the CME front
and/or the methods themselves in ways that are in favor of making
arrival time predictions but difficult to assess quantitatively
without detailed modeling of the global heliosphere.

2. A proper treatment of CME geometry must be performed in
determining CME Sun-to-Earth kinematics, especially when the CME
propagation direction is far away from the observer. The HM
approximation may overestimate the size of CMEs near the Sun, and
the F$\beta$ assumption of a relatively compact structure may become
invalid at large distances due to CME expansion. In the current
cases the two STEREO spacecraft were observing the Earth-directed
CMEs from behind the Sun. While the effect that the two spacecraft
see different parts of a CME gives rise to errors in the
triangulation results, the restrictions of the assumed CME
geometries also contribute. The F$\beta$ geometry gives a distance
\citep[see][and references therein]{liu10b}
$$r_{F\beta} = \frac{d\sin\alpha}{\sin(\alpha+\beta)},$$
where $d$ is the distance of the spacecraft from the Sun, $\alpha$
is the elongation angle of the feature being tracked, and $\beta$ is
the propagation angle of the feature relative to the Sun-observer
line. When the propagation direction is far away from the observer
(say $\pi/2<\beta<\pi$), the combination of $\alpha$ and $\beta$ is
such that, as $\alpha$ increases, the denominator in the above
equation decreases much faster than the increase in the numerator.
This could lead to an apparent late acceleration as we see in the
current cases. The HM geometry results in a distance \citep[see][and
references therein]{liu10b}
$$r_{HM} = \frac{2d\sin\alpha}{1+\sin(\alpha+\beta)}.$$
As can be seen from this equation, the HM approximation can reduce
the late acceleration by introducing weights in the denominator and
numerator. In the 2012 March 7 case we see a slight unphysical late
acceleration from the HM approximation. The situation seems worse
when the above two equations are applied to track fitting, i.e.,
single spacecraft analysis \citep[e.g.,][]{harrison12, rollett12}. A
track fitting approach assuming the HM geometry overestimates the
near-Earth speeds of the 2012 January 23 and March 7 CMEs by
600-1000 km s$^{-1}$ (C. M\"{o}stl 2013, private communication).

The above discussion indicates a warning for interpreting and
forecasting far-sided events. As STEREO spends about 8 years behind
the Sun, many Earth-directed CMEs will be observed by STEREO as
back-sided events. Great caution should be taken when using the CME
geometries (in both track fitting and triangulation) to interpret
the imaging observations and forecast the arrival and speed of CMEs
at the Earth. It is appealing to suggest an intermediate geometry
between the two approaches, e.g., the self-similar expansion (SSE)
model for which the F$\beta$ and HM geometries are limiting cases
\citep{davies12, mostl12}. An attempt to use the SSE geometry for
triangulation is being made \citep{davies13}. This intermediate
geometry could help solve some of the problems. Note that, however,
even the HM geometry overestimates the speed observed at the Earth.
This likely results from the fact that the spherical front
approximation becomes progressively worse, due to flattening of the
front as it interacts with the ambient solar wind. The SSE model, as
an intermediate geometry, probably still overestimates the speed at
1 AU. Also note that for back-sided CMEs the propagation direction
and speed depend ``even more than usual" on the assumed radius of
curvature: the CME nose is almost never observed by the two
spacecraft (only the flanks are), and a different radius of
curvature than what is assumed could make a big difference in the
propagation direction and speed at large distances.

3. Our approach to comparing wide-angle heliospheric imaging
observations with interplanetary radio type II bursts provides a
novel tool in investigating CME propagation characteristics. We
believe that the assessment of CME kinematics by comparing the
imaging observations with type II bursts over the entire Sun-Earth
distance is the first demonstration of this kind of analysis. It
fulfills an objective that STEREO was designed to achieve. In
general, the radio observations are not as widely used as the images
because their integration with other measurements requires special expertise. The novel
approach we use provides a means with which to integrate the imaging
and radio data. It may lead to greater use of those combined data
sets in event analysis, such as to probe the nature of the radio
source region.

This merged imaging, radio and in situ study further demonstrates
the triangulation concept for future CME observations and space
weather forecasting initially proposed by \citet{liu10b}. This
triangulation concept (tentatively called REal-time Sun-earth
Connections Observatory, or RESCO in short) places dedicated
spacecraft at L4 and L5 to make routine remote sensing and in situ
observations. L4 and L5 co-move with the Earth around the Sun, which
enables a continuous monitoring of the Sun and the whole space along
the Sun-Earth line. Triangulation with two such spacecraft makes it
possible to unambiguously derive the true path and velocity of CMEs
in real time. This would be extremely important for practical space
weather forecasting as well as CME Sun-to-Earth research. This work
also provides scientific guidance for spacecraft missions that
intend to look at CME propagation in the ecliptic plane from high
latitude, e.g., the POLAR Investigation of the Sun
\citep[POLARIS;][]{appourchaux09} and the Solar Polar ORbit
Telescope \citep[SPORT;][]{wu11}.

\acknowledgments The research was supported by the STEREO project
under grant NAS5-03131, by the SPORT project under grant Y129164CBS,
by the Recruitment Program of Global Experts of China under grant
Y3B0Z1A840, by the Specialized Research Fund for State Key
Laboratories of China, and by the CAS/SAFEA International
Partnership Program for Creative Research Teams. N. Lugaz was
supported by NSF grant AGS1239704. C. M\"{o}stl has received funding
from the European Union Seventh Framework Programme (FP7/2007-2013)
under grant agreement nr. 263252 [COMESEP], and was supported in
part by a Marie Curie International Outgoing Fellowship within the
7th European Community Framework Programme. We acknowledge the use
of data from STEREO, Wind and GOES. This article commemorates our
friend and coauthor, R. P. Lin, who passed away during the study.

\clearpage

\begin{figure}
\centerline{\includegraphics[width=20pc]{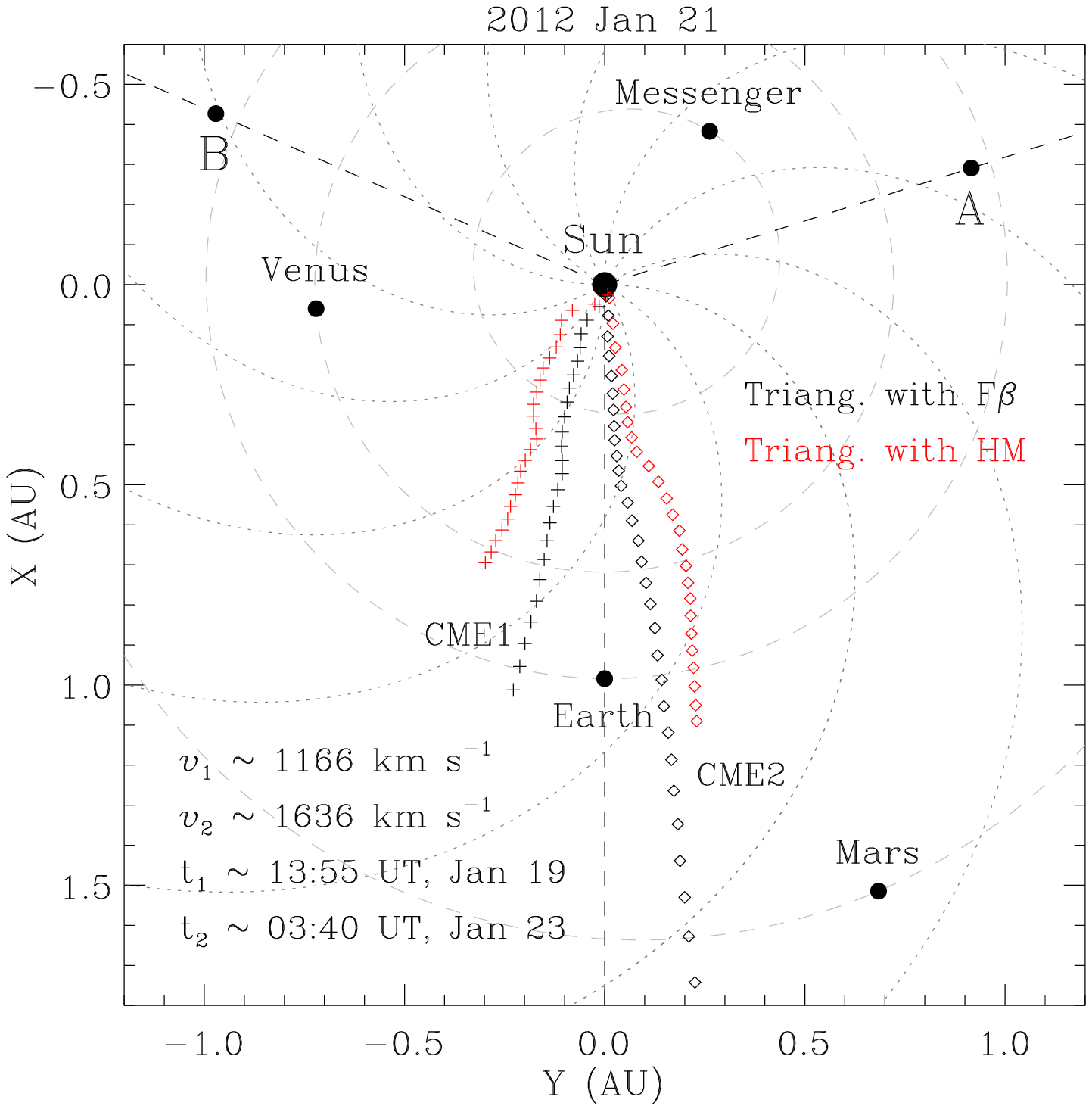}
\includegraphics[width=20pc]{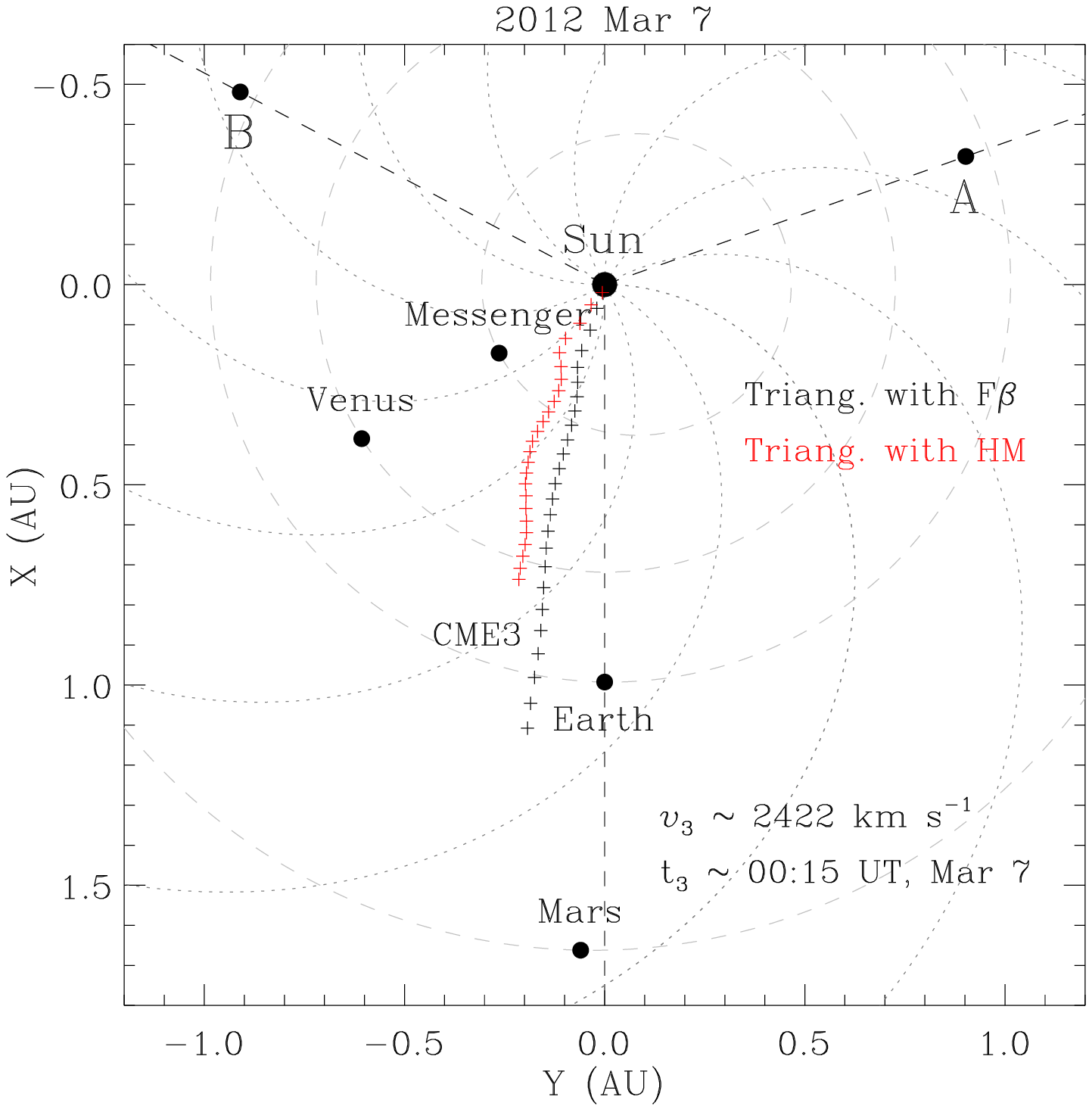}}
\caption{Positions of the spacecraft and planets in the ecliptic
plane on 2012 January 21 (left) and 2012 March 7 (right). The gray
dashed curves indicate the planetary orbits, and the dotted lines
show Parker spiral magnetic fields created with a solar wind speed
of 450 km s$^{-1}$. The trajectories of the CMEs of interest, which
are obtained from triangulation with the F$\beta$ approximation
(black) and triangulation with the HM approximation (red),
respectively, are marked by crosses for CME1 and CME3 and diamonds
for CME2. The estimated CME peak speed and launch time on the Sun
are also given. Note that the two STEREO spacecraft were observing
the CMEs from behind the Sun.}
\end{figure}

\clearpage

\begin{figure}
\centerline{\includegraphics[height=28pc]{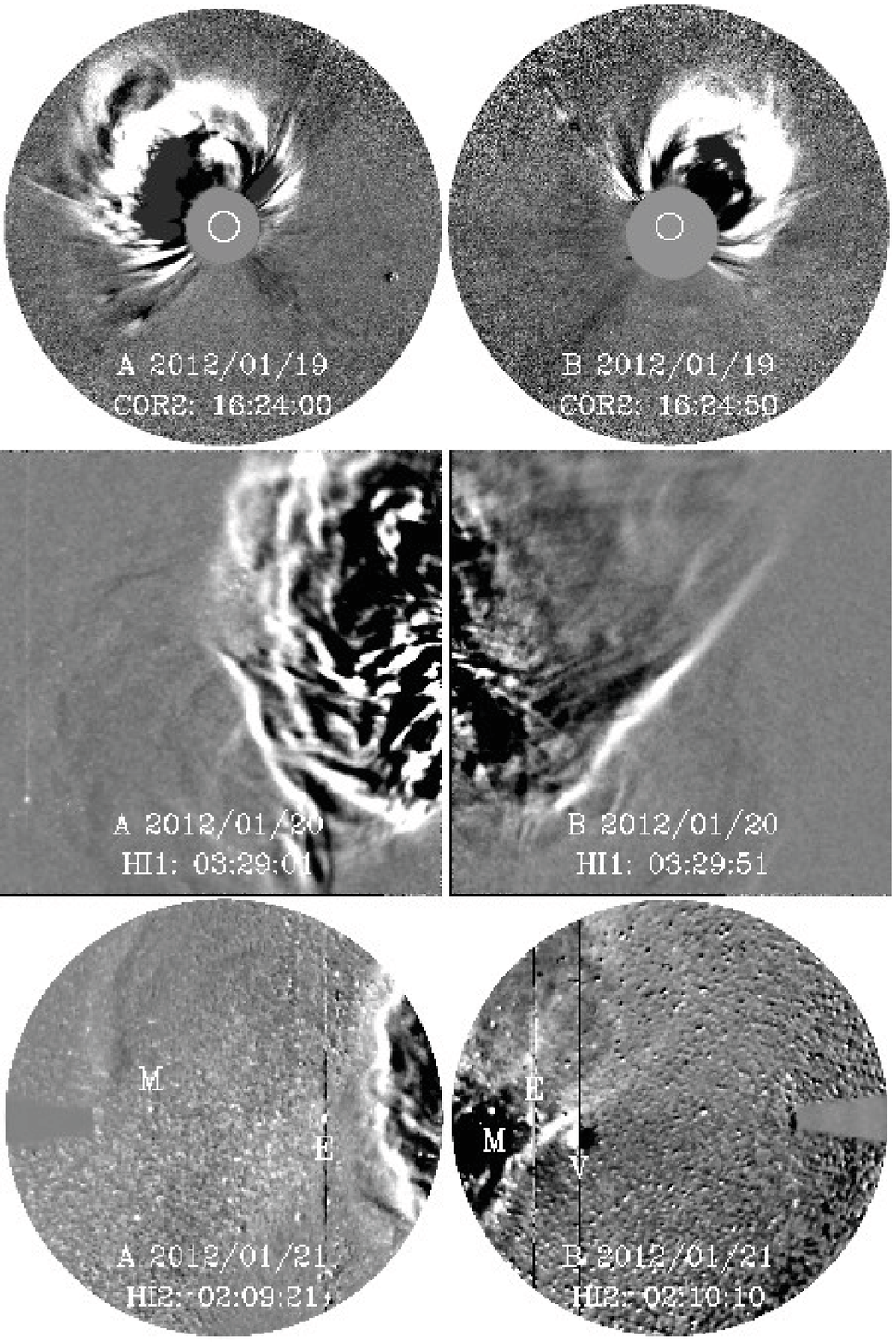}
\hspace{0.5pc}\includegraphics[height=28pc]{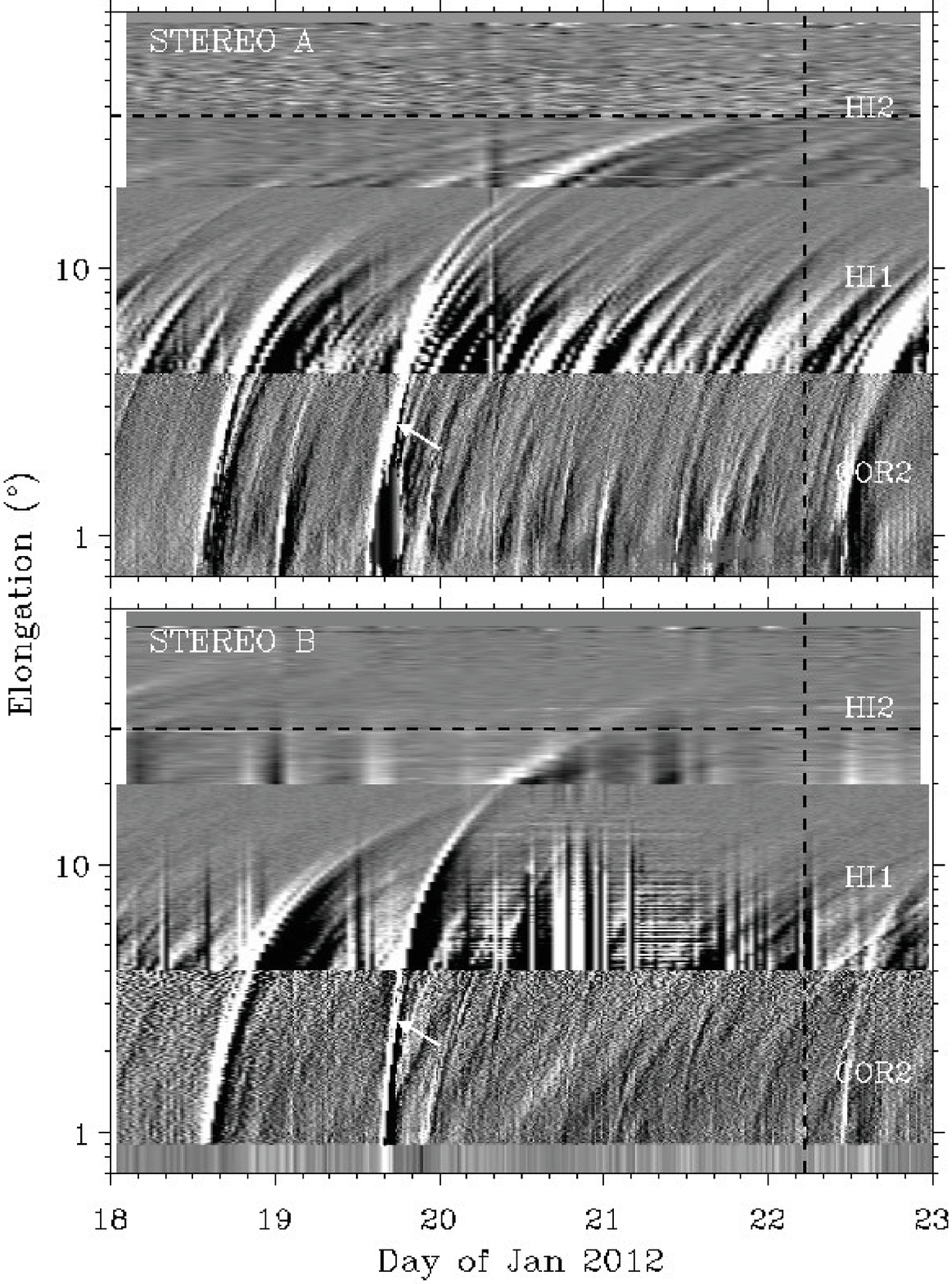}} 
\caption{Left: Evolution of the 2012 January 19 CME
observed by STEREO A and B near simultaneously. From top to bottom,
the panels show the running difference images from COR2, HI1 and
HI2, respectively. The positions of the Earth (E), Mars (M) and
Venus (V) are labeled in the HI2 images. Note a preceding CME
visible to COR2 of STEREO A. Right: Time-elongation maps constructed
from running difference images of COR2, HI1 and HI2 along the
ecliptic plane for STEREO A (upper) and B (lower). The arrow
indicates the track associated with the CME of interest. The
vertical dashed line shows the observed arrival time of the
CME-driven shock at the Earth, and the horizontal dashed line marks
the elongation angle of the Earth.}
\end{figure}

\clearpage

\begin{figure}
\epsscale{0.75} \plotone{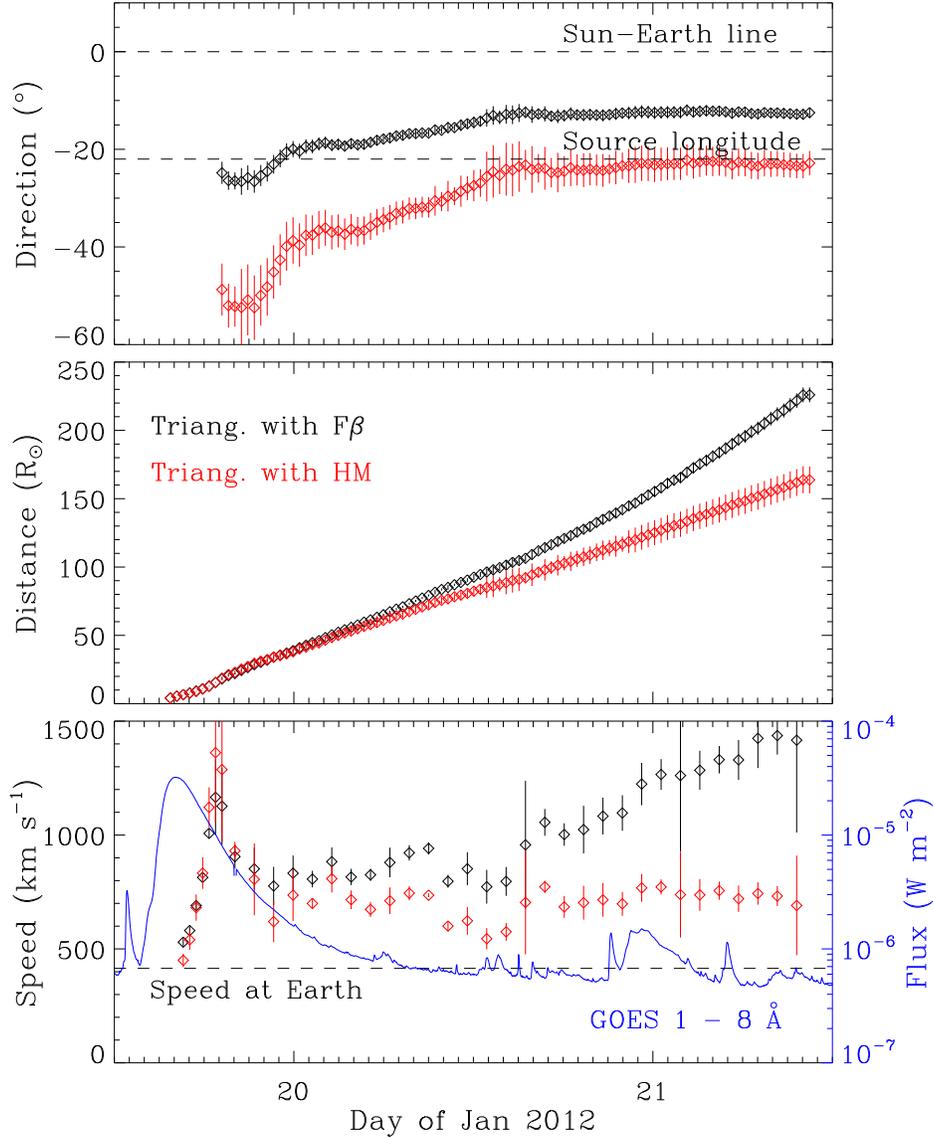} \caption{Propagation direction,
radial distance and speed profiles of the leading edge of the 2012
January 19 CME derived from triangulation with F$\beta$ (black) and
HM (red) approximations. The Sun-Earth line and the longitude of the
CME source location on the Sun are indicated by the dashed lines in
the top panel. The speeds are calculated from adjacent distances
using a numerical differentiation with three-point Lagrangian
interpolation, and then every three points after the acceleration
phase are binned, so shown after the acceleration phase are average
values and standard deviations within the bins. The dashed line in
the bottom panel marks the average solar wind speed in the sheath
region behind the shock observed in situ near the Earth. Overlaid on
the speeds is the GOES X-ray flux (scaled by the blue axis).}
\end{figure}

\clearpage

\begin{figure}
\epsscale{0.8} \plotone{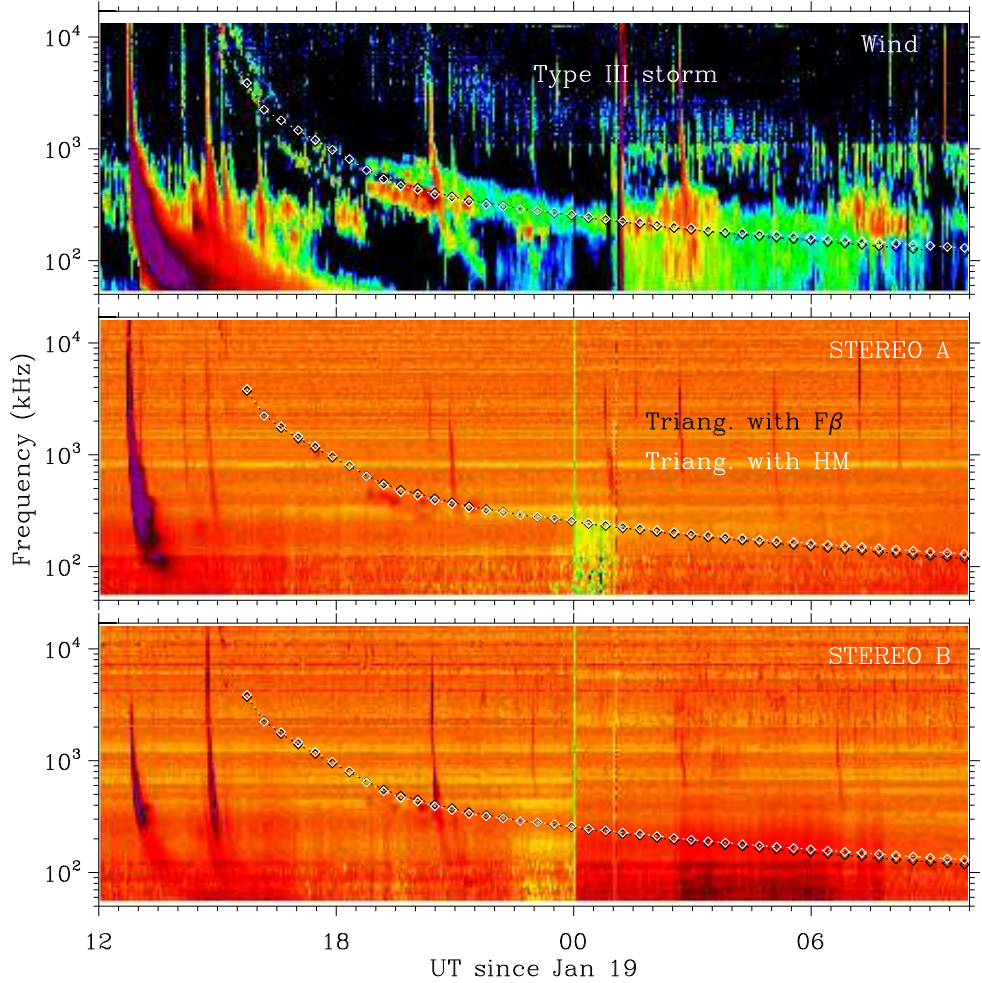} \caption{Dynamic spectrum (color
shading) associated with the 2012 January 19 CME from Wind, STEREO A
and B. The CME leading edge distances derived from triangulation
with F$\beta$ (black) and HM (white) approximations are converted to
frequencies by using the Leblanc density model, which are then
plotted over the dynamic spectrum. Note a type III storm observed by
Wind, which seems to have an overall drift rate similar to that of
the type II burst.}
\end{figure}

\clearpage

\begin{figure}
\epsscale{0.75} \plotone{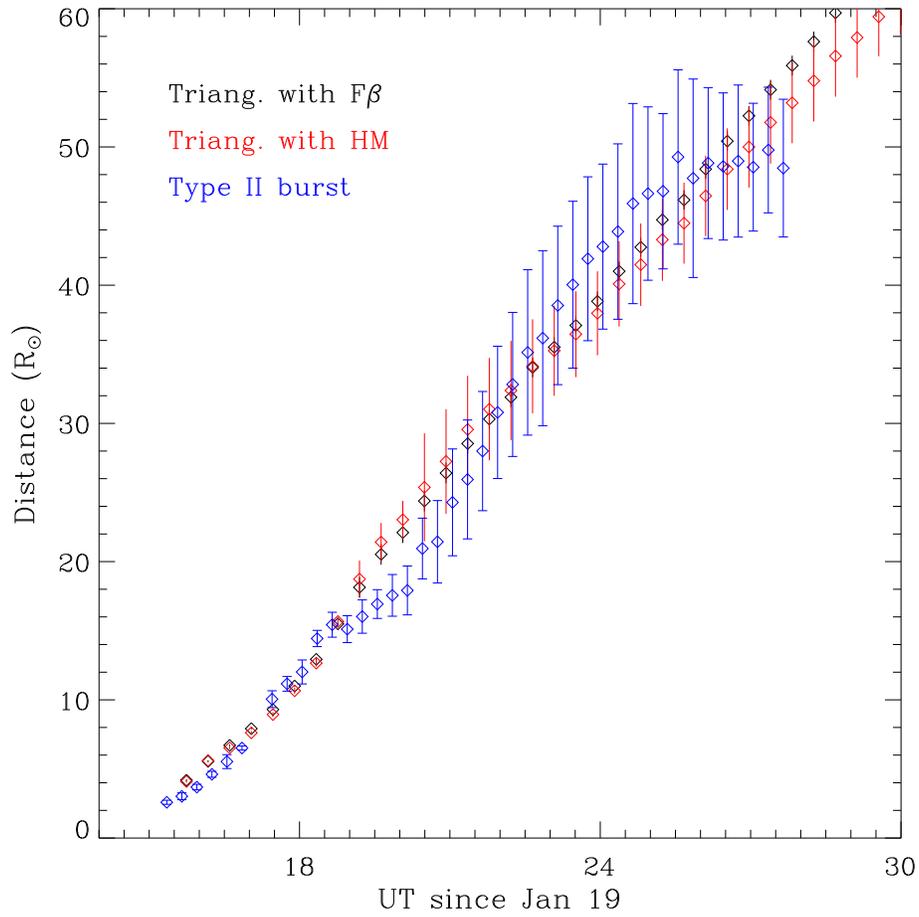} \caption{Comparison between the
CME leading edge distances derived from triangulation with F$\beta$
(black), triangulation with HM (red), and the radio type II burst
(blue).}
\end{figure}

\clearpage

\begin{figure}
\epsscale{0.75} \plotone{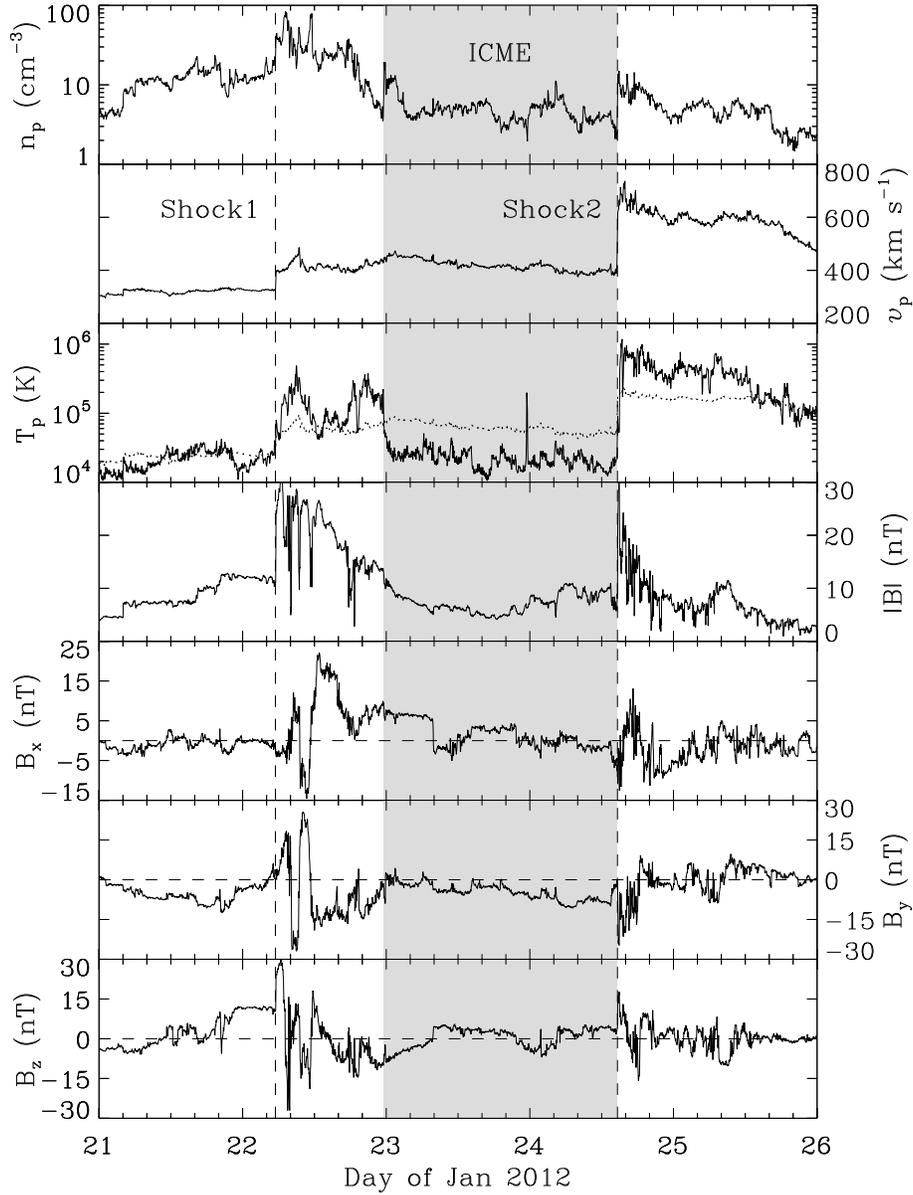} \caption{Solar wind plasma and
magnetic field parameters across the ICME observed at Wind. From top
to bottom, the panels show the proton density, bulk speed, proton
temperature, and magnetic field strength and components,
respectively. The dotted line in the third panel denotes the
expected proton temperature from the observed speed. The shaded
region indicates the ICME interval. The vertical dashed lines mark
two shocks, the first one driven by the present ICME (in situ
counterpart of the 2012 January 19 CME) and the second one by the
2012 January 23 CME. Note that the second shock is beginning to
overtake the ICME from behind. The 2012 January 23 CME does not have
in situ signatures at Wind (except the shock).}
\end{figure}

\clearpage

\begin{figure}
\centerline{\includegraphics[height=28pc]{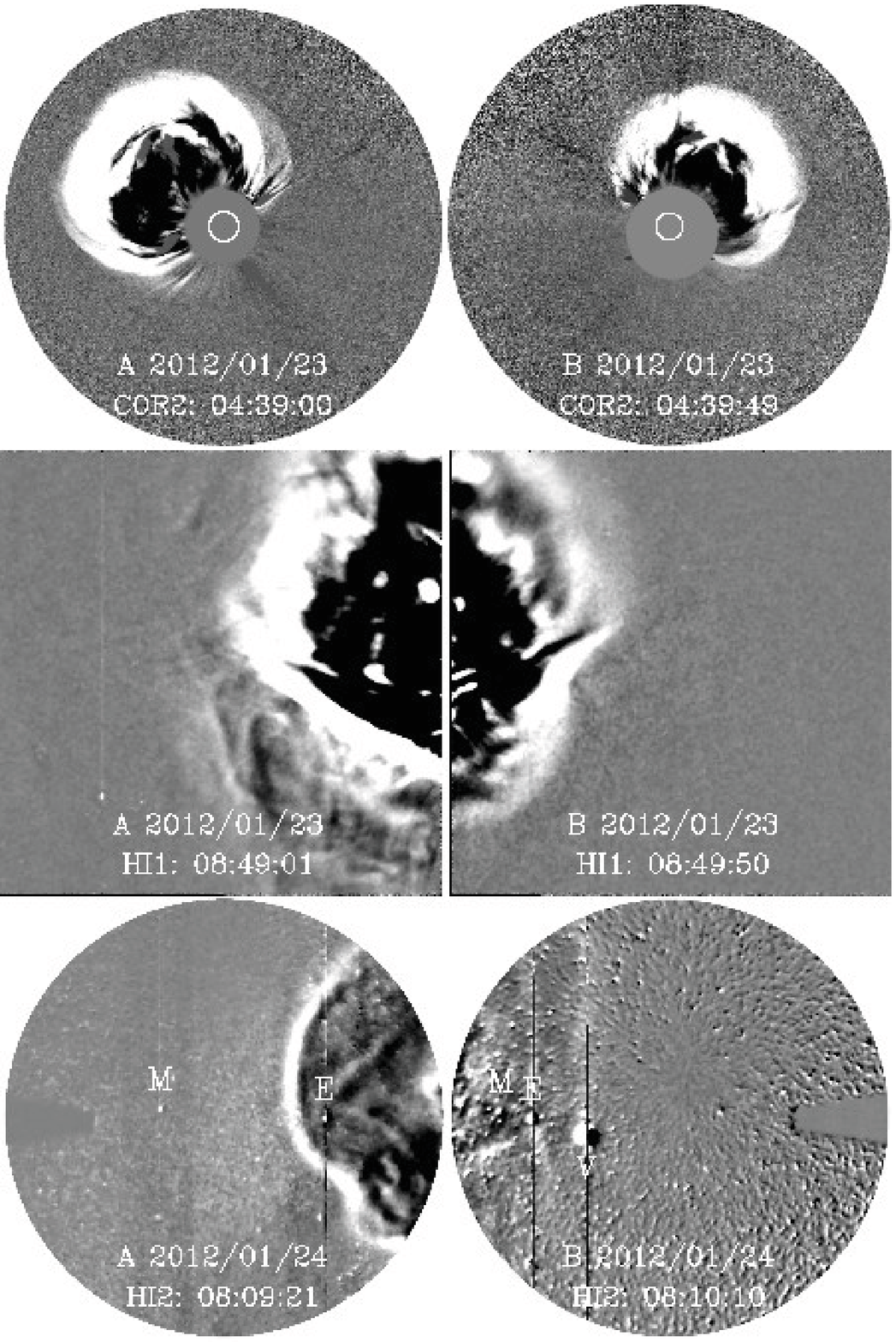}
\hspace{0.5pc}\includegraphics[height=28pc]{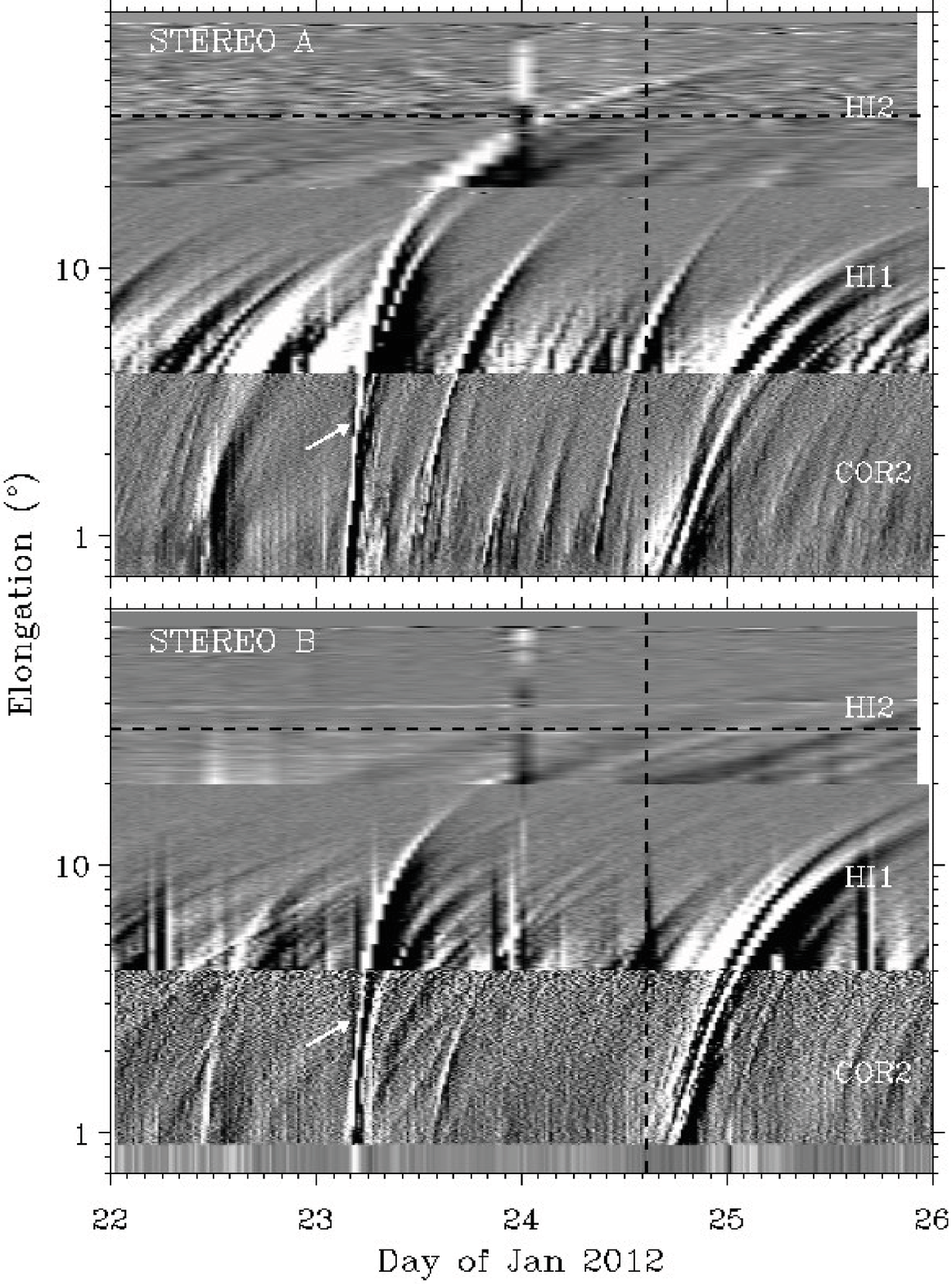}} 
\caption{Similar to Figure~2, but for the 2012
January 23 CME. Note a transition layer around the CME front in
COR2, reminiscent of a shock signature.}
\end{figure}

\clearpage

\begin{figure}
\epsscale{0.75} \plotone{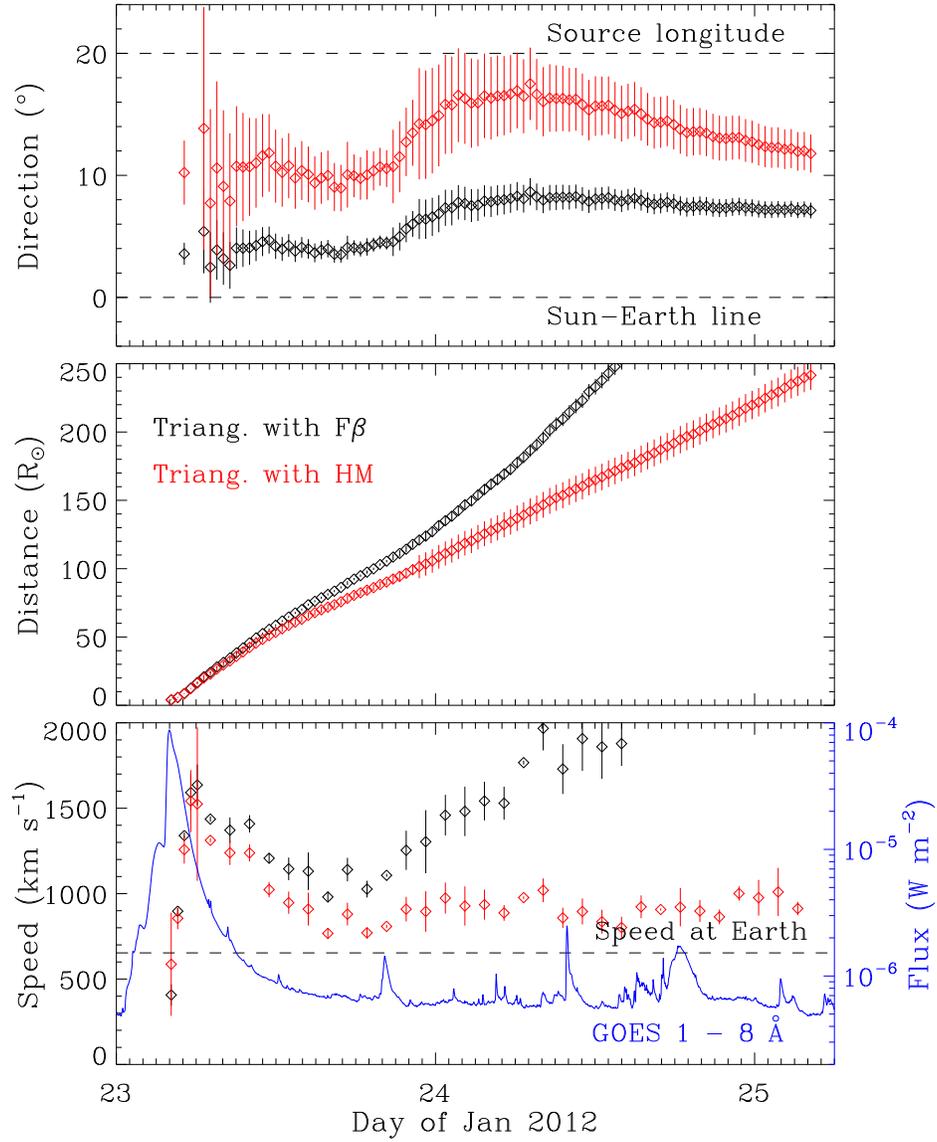} \caption{Similar to Figure~3, but
for the 2012 January 23 CME.}
\end{figure}

\clearpage

\begin{figure}
\epsscale{0.8} \plotone{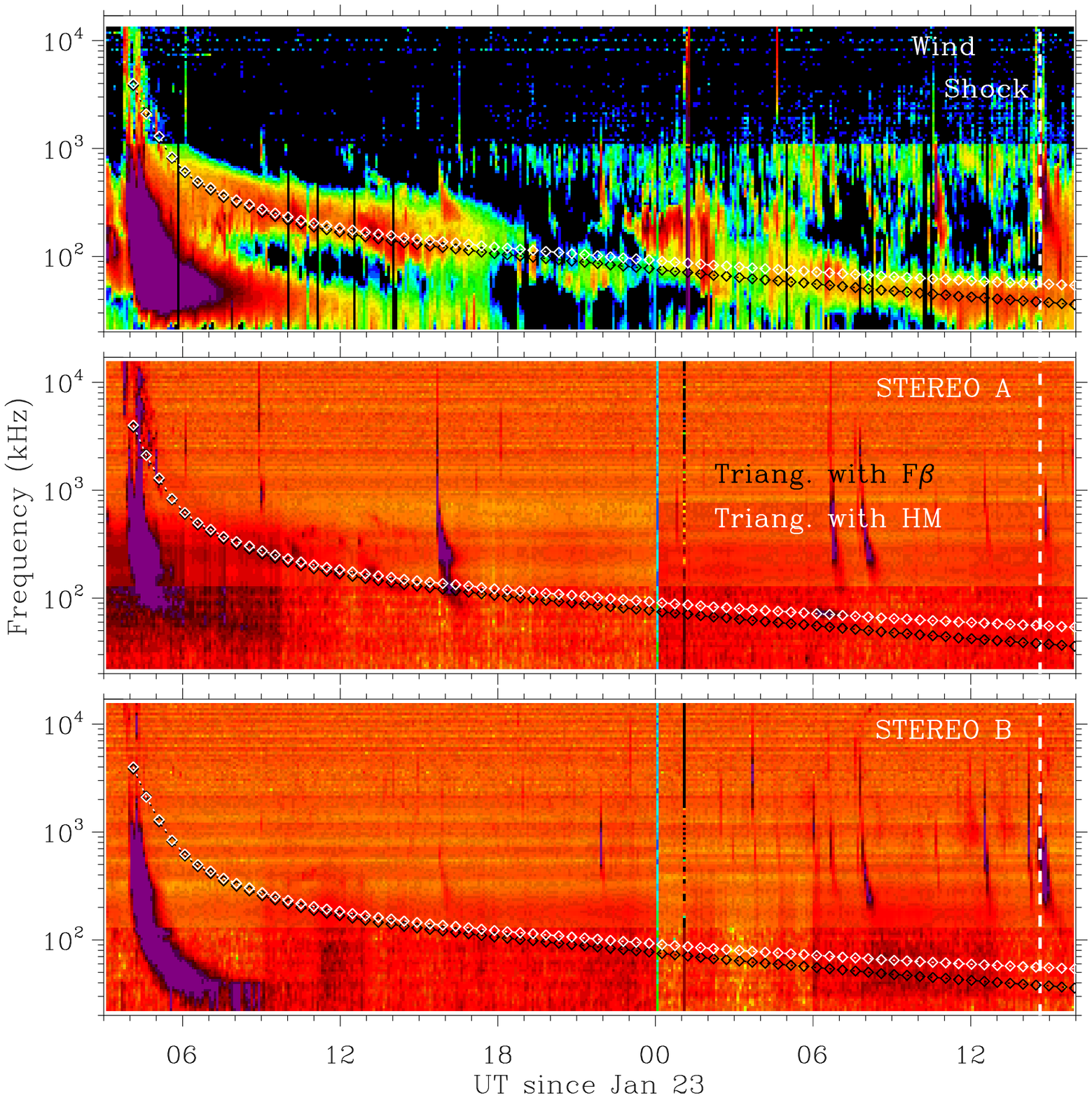} \caption{Similar to Figure~4, but
for the 2012 January 23 event. The vertical dashed line indicates
the observed arrival time of the CME-driven shock at the Earth.}
\end{figure}

\clearpage

\begin{figure}
\epsscale{0.75} \plotone{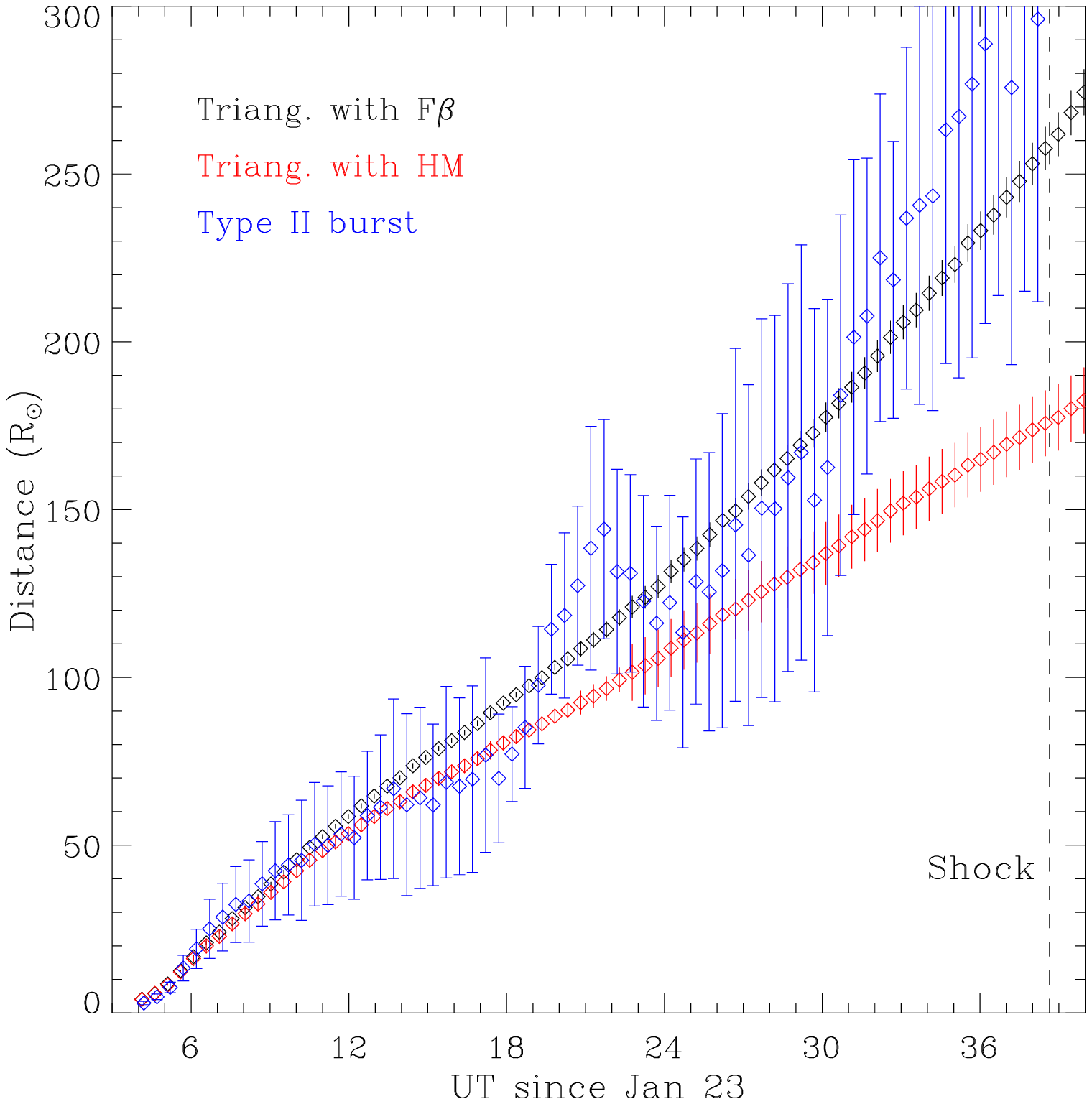} \caption{Similar to Figure~5, but
for the 2012 January 23 CME. The vertical dashed line indicates
the observed arrival time of the CME-driven shock at the Earth.}
\end{figure}

\clearpage

\begin{figure}
\centerline{\includegraphics[height=28pc]{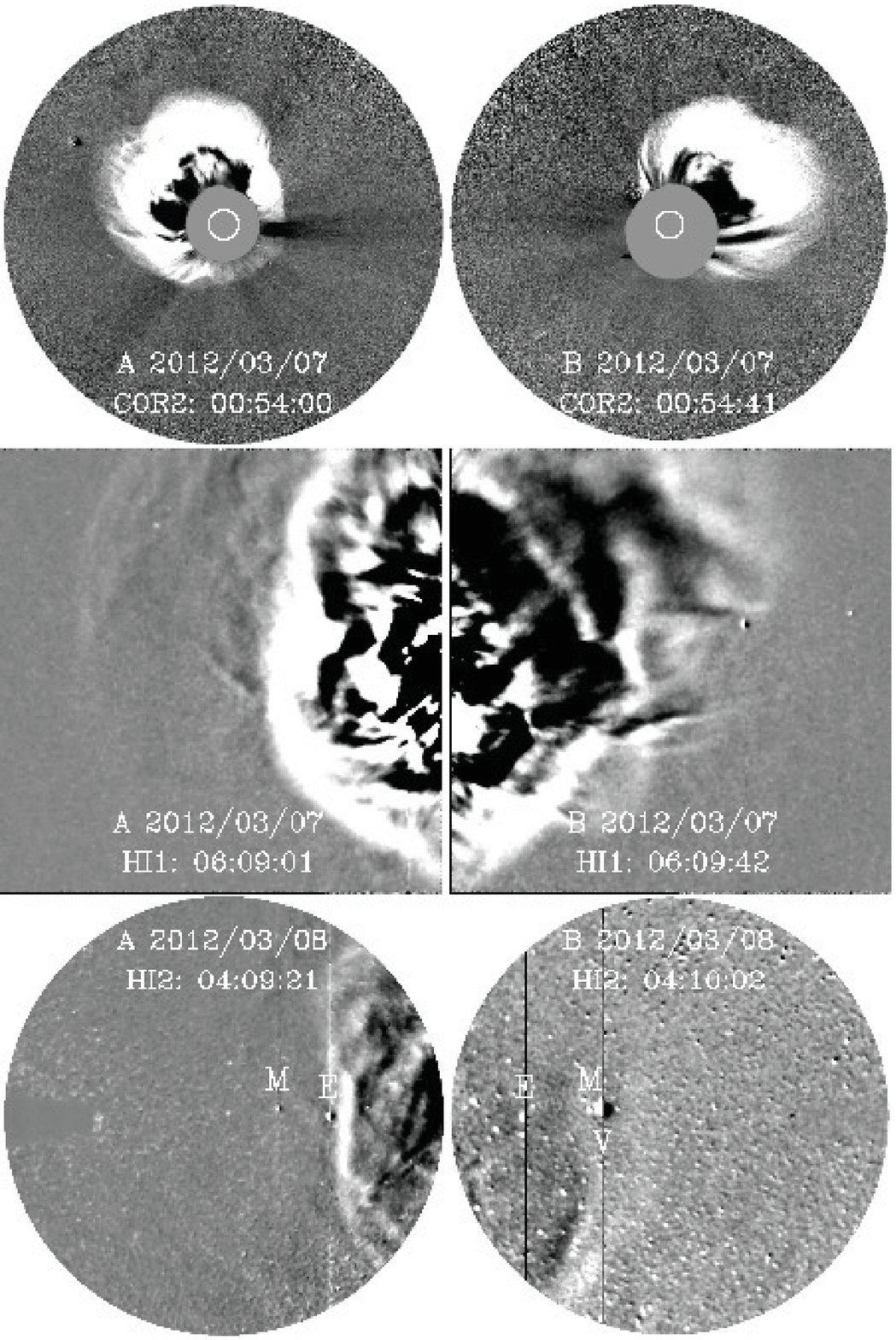}
\hspace{0.5pc}\includegraphics[height=28pc]{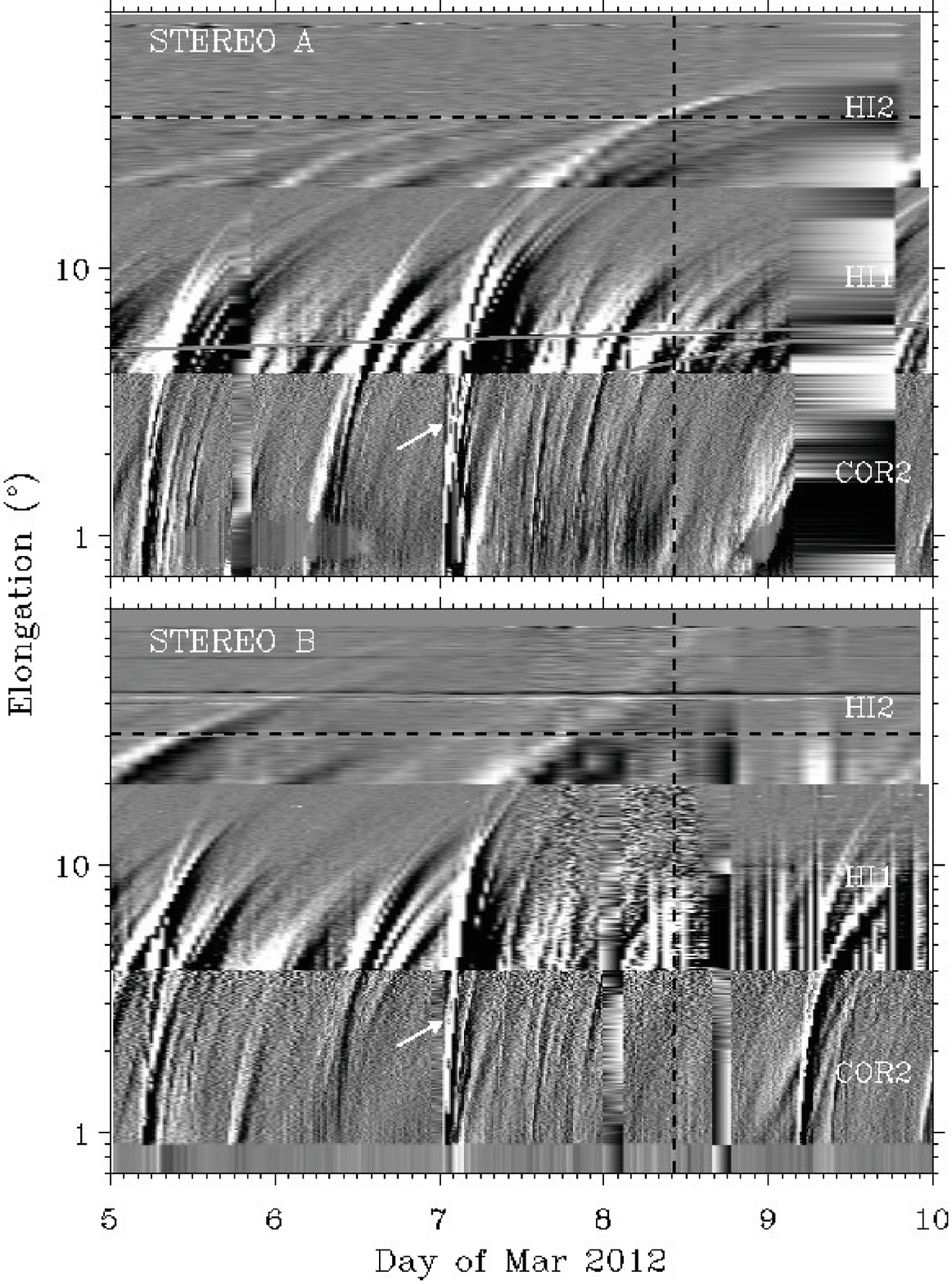}} 
\caption{Similar to Figure~2, but for the 2012 March
7 CME. A weak edge around the CME front is also observed in COR2.}
\end{figure}

\clearpage

\begin{figure}
\epsscale{0.75} \plotone{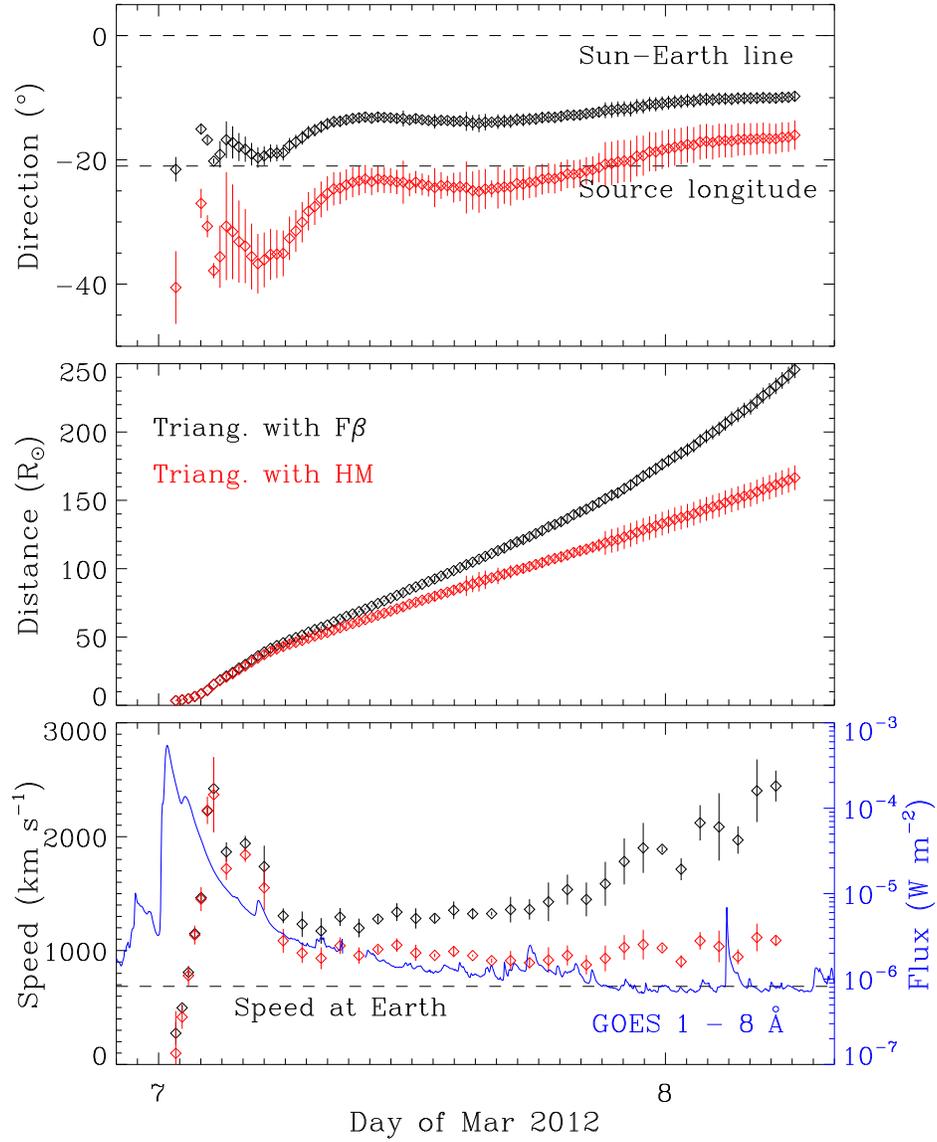} \caption{Similar to Figure~3, but
for the 2012 March 7 CME.}
\end{figure}

\clearpage

\begin{figure}
\epsscale{0.8} \plotone{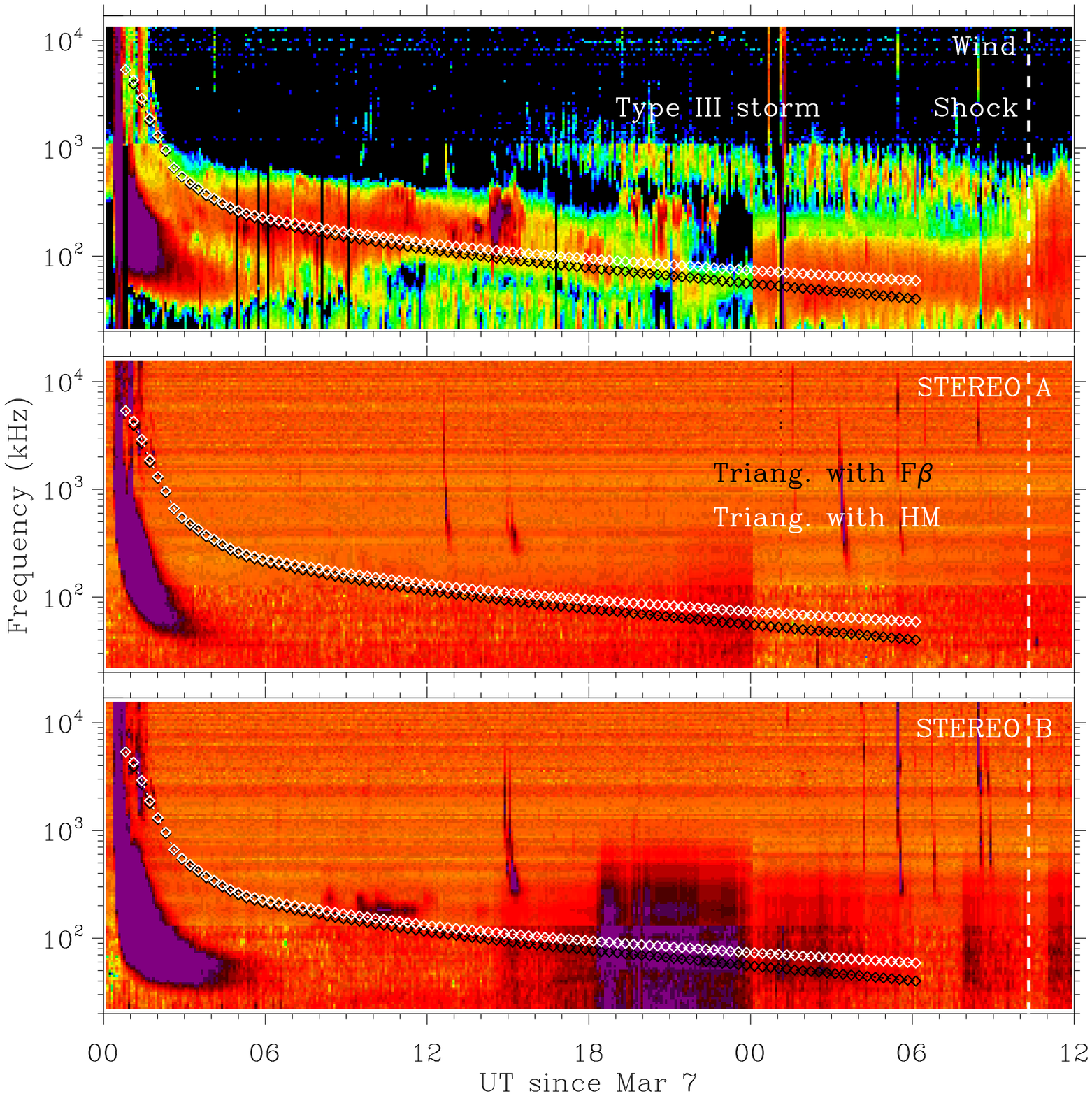} \caption{Similar to Figure~4, but
for the 2012 March 7 event. The vertical dashed line indicates the
observed arrival time of the CME-driven shock at the Earth.}
\end{figure}

\clearpage

\begin{figure}
\epsscale{0.75} \plotone{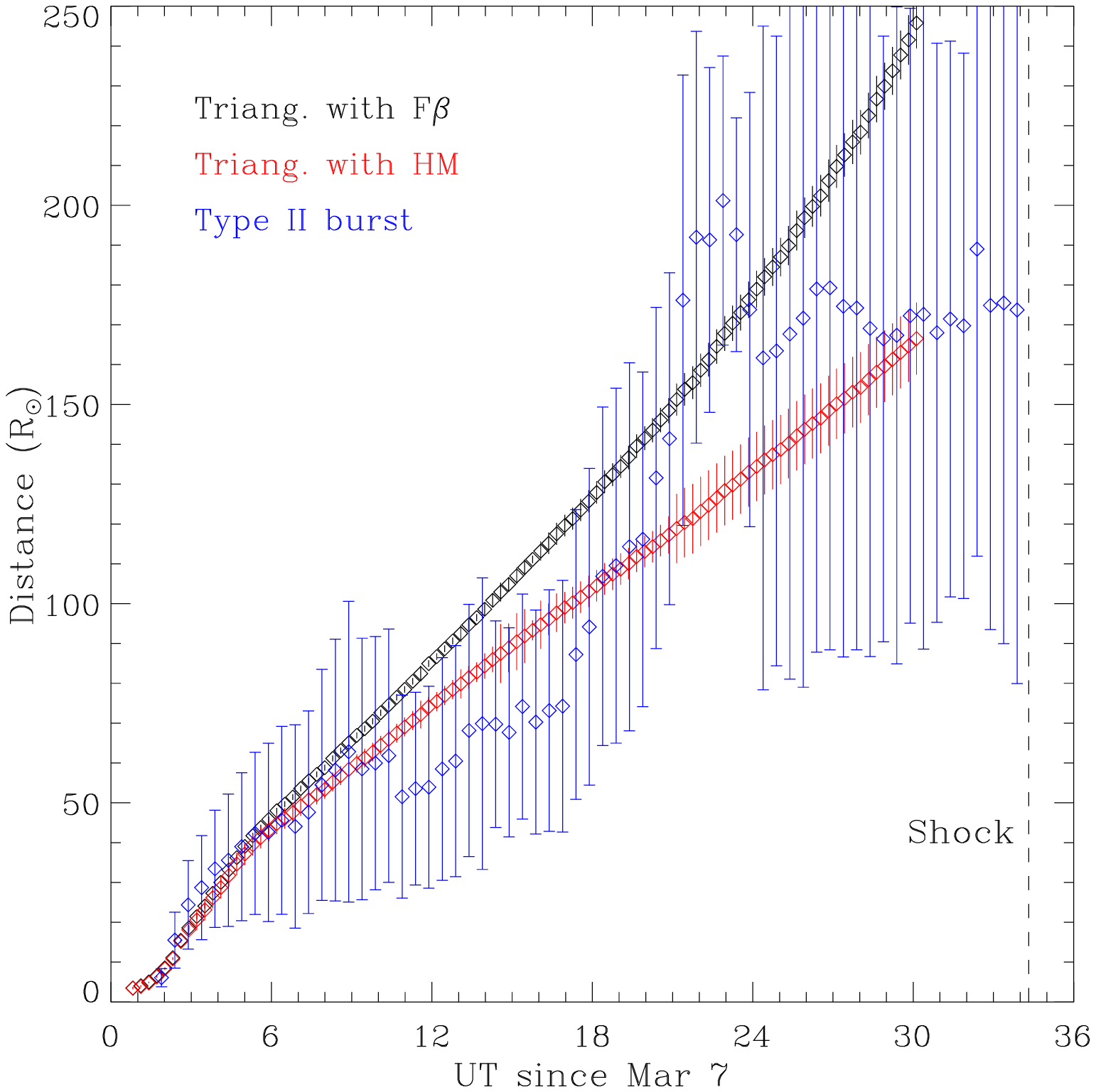} \caption{Similar to Figure~5, but
for the 2012 March 7 CME. The vertical dashed line indicates
the observed arrival time of the CME-driven shock at the Earth.}
\end{figure}

\clearpage

\begin{figure}
\epsscale{0.75} \plotone{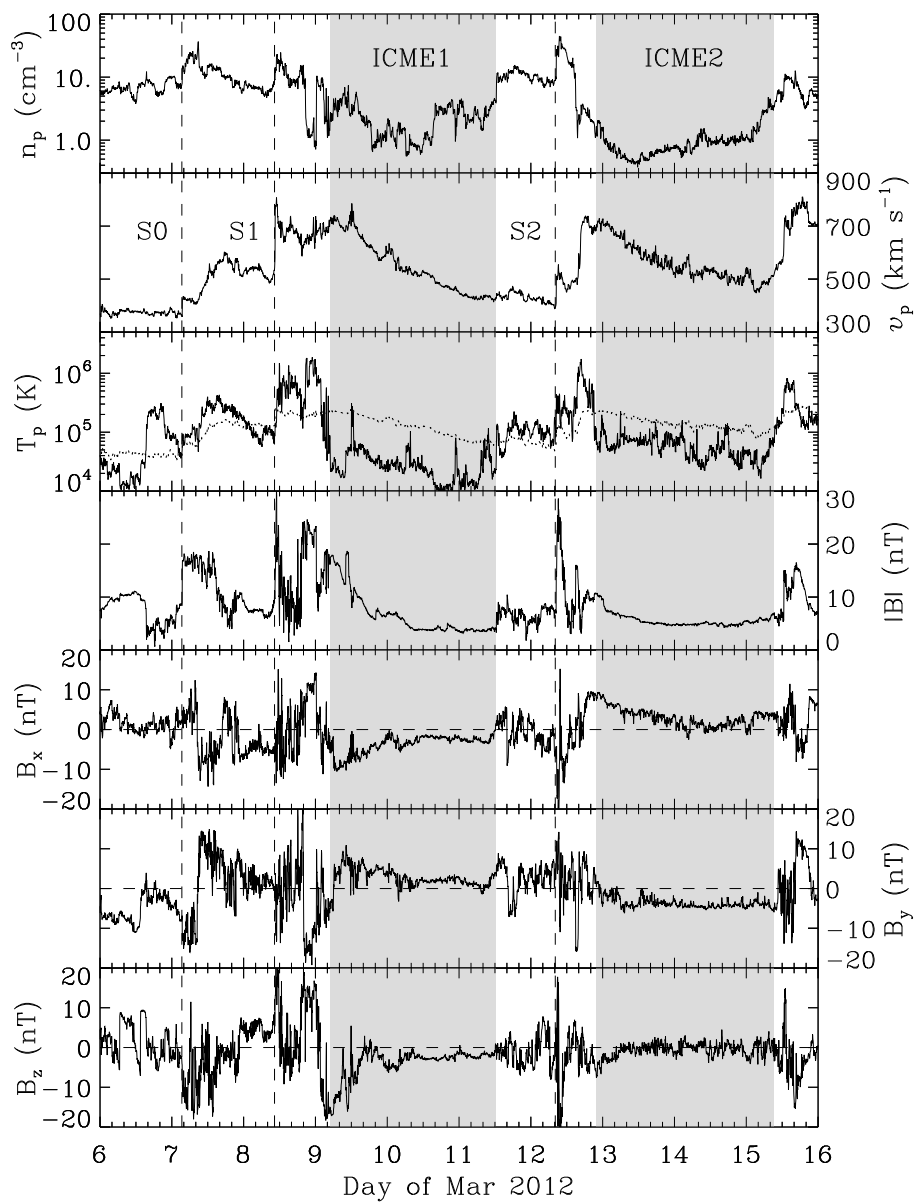} \caption{Similar to Figure~6, but
for the 2012 March 7 event. The first ICME (ICME1) and its preceding
shock (S1) are produced by the March 7 CME, while the second ICME
(ICME2) and its shock (S2) are generated by a CME from March 10. A
weak shock (S0) ahead of S1 seems driven by a CME on March 5, but no
clear ICME signatures are observed at Wind (except the shock).}
\end{figure}

\end{document}